# Zn-doped P-type InAs Nanocrystal Quantum Dots

*Lior Asor, Jing Liu, Shuting Xiang, Nir Tessler, Anatoly I. Frenkel\*, Uri Banin\**

L. Asor, Prof. U. Banin
The Institute of Chemistry and The Center for Nanoscience and Nanotechnology
The Hebrew University of Jerusalem
Jerusalem 91904, Israel

Prof. J. Liu
Physics Department
Manhattan College
Riverdale, New York 10471, USA

S. Xiang, Prof. A. I. Frenkel
Department of Materials Science and Chemical Engineering
Stony Brook University
Stony Brook, New York 11794, USA

Chemistry Division
Brookhaven National Laboratory
Upton, New York 11973, USA

Prof. N. Tessler
The Zisapel Nano-Electronics Center
Department of Electrical Engineering
Technion – Israel Institute of Technology
Haifa 32000, Israel

**ABSTRACT**

Doped heavy metal-free III – V semiconductor nanocrystal quantum dots are of great interest both from the fundamental aspects of doping in highly confined structures, and from the applicative side of utilizing such building blocks in the fabrication of p-n homojunction devices. InAs nanocrystals, that are of particular relevance for short wave IR detection and emission applications, manifest heavy n-type character poising a challenge for their transition to p-type behavior. We present p-type doping of InAs nanocrystals with Zn – enabling control over the charge carrier type in InAs QDs field effect transistors. The post-synthesis doping reaction mechanism is studied for Zn precursors with varying reactivity. Successful p-type doping was achieved by the more reactive precursor, diethylzinc. Substitutional doping by $Zn^{2+}$ replacing $In^{3+}$ is established by X-ray absorption spectroscopy analysis. Furthermore, enhanced near IR photoluminescence is observed due to surface passivation by Zn as indicated from elemental mapping utilizing high resolution electron microscopy corroborated by X-ray photoelectron spectroscopy study. The demonstrated ability to control the carrier type, along with the improved emission characteristics, paves the way towards fabrication of optoelectronic devices active in the short wave IR region utilizing heavy-metal free nanocrystal building blocks.

# 1. Introduction

The field of semiconductor colloidal quantum dots (CQDs)-based optoelectronics draws attention worldwide, as new technologies call for high volume, cost-effective fabrication of optoelectronic devices with precise control over the optical properties and complementary metal-oxide-semiconductor (CMOS) compatibility.[1,2] Of specific interest are QDs with tunable band gap across the near infrared and short-wave infrared (SWIR) spectra for applications in IR detection and vision[3,4], photovoltaics,[5–7] and telecommunications.[8–11] Numerous efforts and focus have been given to Pb- and Hg- based semiconductor QDs, owing to the well-established synthetic procedures and well-studied optical properties, perfected over the years.[12–16] However, the implementation of such materials in commercial optoelectronics is restricted due to environmental concerns and regulatory aspects as dictated by the Restriction of Hazardous Substances Directive (RoHS).[17] This stimulates the research and development of novel alternatives, where III–V colloidal QDs such as InAs are considered as a promising candidate material system for infrared optoelectronic applications.[8,18] A particular challenge in InAs QDs, and also for other colloidal QDs, is the control and understanding of doping, and its utilization in optoelectronic devices.[19–23]

The study of III-V colloidal QDs based optoelectronics, and InAs in particular, is hindered, mainly due to synthetic issues, including precursors toxicity and limited control over the size and uniformity of the QDs. In the last few years, several synthetic advancements presented improved tunability of the band gap across the NIR and SWIR.[24–27] Controlled growth of InAs nanocrystals by continuously supplying InAs nanoclusters, enables the synthesis of highly crystalline and monodisperse nanocrystals with narrow absorption features across the near infrared.[28,29] Additional successive growth cycles provide the ability to grow even larger InAs nanocrystals, pushing the absorption feature towards 1600 nm while maintaining a narrow size distribution.[30]

InAs QDs, as synthesized, present n-type character.[22,31] Thus, combining such n-type InAs QDs with another p-type material, allowed to form a heterojunction as was achieved for photovoltaic[29] and IR detection applications.[32–34] However, ideally, one should design a p-n homojunction, where the n- and p-type layers are composed of the same base QD material – as realized before with PbS QDs.[35] Therefore, precise control over the electrical properties of III-V semiconductor QDs, with the ability to produce robust p-type doping is crucial for the future adoption of III-V semiconductors in all-colloidal QDs based optoelectronic devices and the fabrication of InAs QDs-based p-n homojunction to compete with Pb and Hg containing counterparts. Moreover, the study and development of doped III-V colloidal QDs is interesting also from the fundamental aspects of doping in highly confined structure. As such, it is necessary to study and understand the mechanism, structural, and physical properties of the doped system.

We reported previously on an approach to dope pre-synthesized InAs QDs with metal impurities which brings them to the heavy doping regime.[19–21] An advantage of such post-synthesis doping reaction is the ability to separate the control of the QD size and its related initial optoelectronic characteristics, from the doping step, thus emulating the merits of the powerful and widely utilized doping processes in bulk semiconductors. This also enables to directly compare the doped versus pristine nanocrystals made from the same batch.

The effect of the post-synthesis doping approach on the performance of colloidal InAs QDs-based field effect transistors (FETs) established the ability to manipulate the doping level and the majority carrier type. InAs QDs that were post-synthetically doped with Cu presented enhanced n-type doping properties in QDs-based FETs compared to as-synthesized, undoped QDs,[22] in line with previous spectroscopic and theoretical studies.[19,20] Recently, we also developed a modified post-synthetic doping procedure of InAs QDs with Cd, achieving robust p-type doping of InAs QDs in FETs.[23] Using X-ray absorption fine structure (XAFS) spectroscopy measurements, the site-specific location of Cd within the QDs host lattice was identified, showing that it acts as a substitutional dopant to Indium near the surface of the QDs. We also found that surface Cd not only acts as a p-dopant, but also protects the QDs against oxidation, allowing for stable device operation when exposed to air. However, while Cd-doped InAs serves as an excellent model system to study p-type doping effects in colloidal InAs QDs - implementation of such doped QDs is hindered because of the restrictions on the commercial use of Cd as defined in the RoHS.

With this in mind, we sought to develop a doping procedure using a less toxic and RoHS compliant p-type dopant for InAs QDs. Zn is known to induce p-type doping in III-V semiconductor nanowires,[36–39] thus making it a promising candidate. The incorporation of Zn during the synthesis of colloidal InP QDs has been extensively studied in terms of its positive effect on the PL quantum yield.[40–43] Reports are mixed with regards to the Zn location upon such reaction, between surface deposition to some incorporation inside the QD lattice forming an In(Zn)P alloy composition.[43] Use of less reactive Zn-carboxylate precursor was found to mostly lead to surface deposition, with some incorporation of Zn into the lattice while using more reactive shorter chain carboxylate. Recently, $ZnCl_2$ mediated synthesis of small InAs QDs utilizing aminoarsine precursor was reported, yielding improved size distribution and enhanced fluorescence at 860 nm.[27] In this case, the Zn was deposited on the surface without incorporation into the lattice. While the possible formation of impurity states related to Zn in In(Zn)P NCs was studied theoretically,[44] direct observation of consequential p-type doping in Zn containing III-V NC FETs is lacking.

Herein, we introduce a *post-synthesis* doping reaction of InAs QDs with Zn impurities, aiming at RoHS compliant p-type colloidal QDs. To investigate the effect of the precursors chemistry on the resultant doped QDs, two Zn precursors with different reactivity, Zn(Oleate)$_2$ and diethylzinc, were studied. Only the latter, more reactive precursor, yielded successful p-type doping and switching of the majority carrier type from electrons to holes as presented in doped colloidal InAs QDs FETs. For both precursors, the PL is enhanced and more so for the more reactive diethylzinc. To investigate the structural origin of the observed difference in properties, we analyzed the local structure around Zn dopants in the materials prepared from these two precursors using XAFS spectroscopy. XAFS has excellent sensitivity to the local structure of each atomic species (Zn, In and As) that can be used to propose the details of the architecture of the doped QD. Using this method, we have shown in the past that different types of doping scenarios (interstitial, substitutional, segregation, surface doping) were obtained in Cu, Ag, Au, Cd dopants in InAs.[20,21,23,45] In the present work, we aimed at the determination of the Zn location after the doping reaction, hypothesizing that the doped QDs with the p-type character have $Zn^{2+}$ substituting for $In^{3+}$ in the lattice. For the both materials, ultrahigh-resolution scanning transmission electron microscopy (UHR-STEM) with EDS mapping reveals the presence of Zn on the surface, which is also corroborated by X-ray photoelectron spectroscopy (XPS) analysis accompanied by analytical determination of the overall Zn content. XAFS analysis directly manifested the origin of the observed differences in the electronic properties: using the diethylzinc precursor resulted in substitutional doping of InAs, while the Zn(oleate)$_2$ precursor led to the formation of surface Zn oxides. The introduced route to form p-type RoHS compliant colloidal InAs QDs introduces important building block towards the future design and fabrication of an all nanocrystal-based p-n homojunction as basis for SWIR active optoelectronic devices.

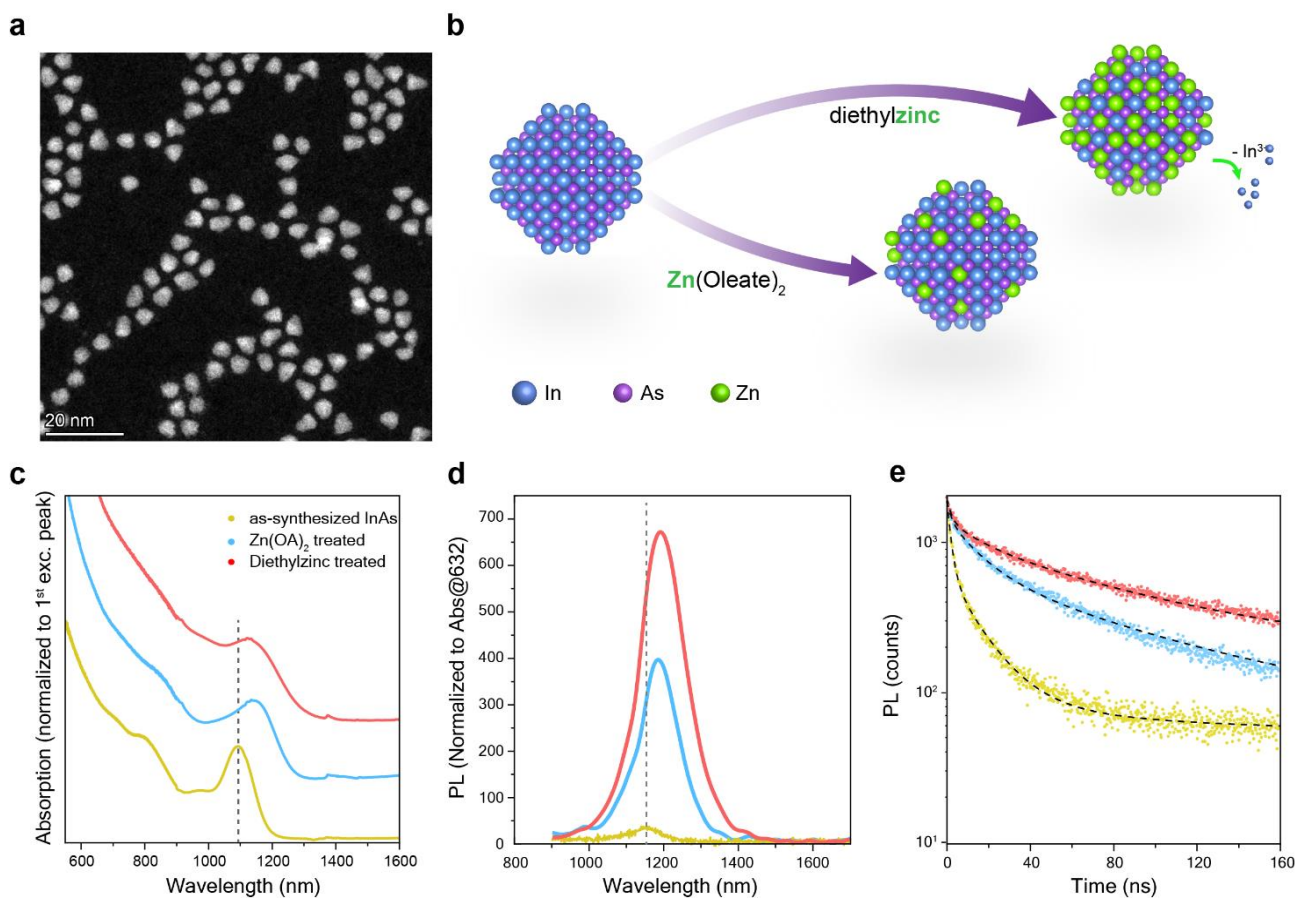

**Figure 1. post-synthetic doping of InAs NCs with Zn**. **a**, STEM HAADF image of the as-synthesized InAs NCs. **b**, Schematic illustration of the post-synthetic doping reaction with either Zn(Oleate)$_2$ or diethylzinc. **c-d**, Absorption and photoluminescence spectra of the as-synthesized and Zn-doped NCs. **e**, Time-resolved PL lifetime measurements of the as-synthesized and Zn-doped InAs NCs.

## 2. Results and Discussion

### 2.1. Synthesis and Doping Reaction of InAs QDs

Colloidal InAs QDs were synthesized utilizing the continuous injection of InAs nanoclusters method with slight modifiactions.[28] This method enables the synthesis of high quality and large scale batches of InAs QDs with tunable band gap between 800 – 1400 nm and offers delicate control over the size distribution of the as-synthesized QDs with narrow absorption and emission features. Briefly, InAs NCs seeds with excitonic absorption peak at 750 nm were synthesized by hot injection reaction between In(Oleate)$_3$ and tris(trimethylsylil)arsine at 300c. Then, pre-synthesized InAs nanoclusters were injected at a constant rate into the reaction flask containing the seeds for continuous, controllable growth of highly monodisperse InAs NCs (further details regarding the synthesis can be found in the supporting information). Figure 1a shows STEM characterization of a monodisperse sample of highly crystalline, as-synthesized 5 nm InAs QDs with trigonal-like shape and a narrow excitonic absorption feature at 1100 nm, indicative of the narrow size distribution.

Post synthesis doping of the as-synthesized InAs QDs with Zn was inspired by our previous work on Cd-doped InAs,[23] using either Zn(Oleate)$_2$ or diethylzinc as Zn precursors of increasing reactivity. Illustrated schematically in Figure 1b, the reaction with Zn was carried out by dropwise addition of either of the two Zn precursors with a nominal Zn:In overall ratio of 1.0:1.0, to a preheated solution of as-synthesized InAs QDs in 1-octadecene and oleylamine at 260°c under inert atmosphere, and the reaction was left for 3 hours. Absorption and photoluminescence (PL) spectra of the QDs reacted with Zn were monitored during the reaction, showing a gradual red-shift of the 1$^{st}$ exciton absorption peak and a significant improvement in PL intensity as Zn addition proceeded for both precursors. The doping reaction was stopped when no further changes in absorption and PL were observed. At the end of the doping reaction with diethylzinc, the doped QDs are purified from the crude solution. The latter was found to contain byproducts of In$_2$O$_3$ particles, as indicated by XRD characterization (see figure S1). This already serves as a first indication for the substitution of Indium with Zinc. Doping with Zn(Oleate)$_2$ on the other hand, did not produce such byproducts.

The absorption and PL spectra of the as-synthesized InAs QDs and the purified Zn-reacted QDs are presented in figure 1c,d. In both Zn-reaction routes, the absorption feature is red-shifted and broadens. However, this broadening is more pronounced for the QDs doped with diethylzinc, the more reactive precursor, implying for successful doping and higher Zn content within the QDs lattice. In the limit of heavy doping, both in bulk semiconductors as well as in nanocrystals, band tailing occurs due to the distortion in the lattice. This leads to red shifted band gap and to broadening as observed previously for heavily doped InAs NCs.[19,22,23] In addition, the PL of the diethylzinc doped QDs increases significantly, by more than an order of magnitude relative to the as-synthesized QDs and a 2-fold increase in the PL quantum yield compared to the Zn(Oleate)$_2$ doped QDs.

The effect of Zn treatment on the PL properties was further characterized using time-resolved photoluminescence lifetime (TRPL) measurements, depicted in figure 1e. The PL decay is elongated upon post-synthesis treatment with Zn with the most significant elongation observed for the diethylzinc case, consistent with its highest PL intensity. By fitting the time-resolved PL decays to a tri-exponential function with fixed lifetime decay constants of 2, 15, and 80 ns, an increase of the long lifetime pre-exponential factor (A3) and a decrease of the short one (A1) (see figure S2) is observed upon Zn treatment. This indicates the role of Zn on the surface for the passivation of trap sites. Such improvement in the PL quantum yield of near IR and SWIR emitting QDs is beneficial also for IR lasing and telecommunications applications, and we demonstrate here that just a simple treatment with Zn, as was also demonstrated in InP QDs,[46,47] has a large impact on the PL quantum yield of colloidal InAs QDs. As herein the focus is on the control over the charge carrier type in InAs QDs, we next describe the effect of Zn doping and Zn-precursor reactivity on the p-type doping efficiency in InAs QDs based FETs.

**2.2. Field Effect Transistors Characterization**

InAs QDs-based FETs were fabricated using our established method reported previously.[22,23] Briefly, a 30 nm thick QDs film was fabricated by spin coating a concentrated InAs QDs solution on top of heavily doped Si substrate covered with 100 nm of SiO$_2$ and additional 10 nm of HfO$_2$. Solid state ligand exchange process was performed to replace the long alkyl chain native ligands with 1,2 Ethanedithiol (EDT) to decrease inter-dot distance and improve conductivity. The device was completed by deposition of Au source/drain interdigitated electrodes.

Figure 2 shows the output and transfer characteristics of as-synthesized and Zn-doped InAs QDs FETs presenting the effect of Zn doping on the electrical characteristics of InAs QDs films. The output characteristics of the as-synthesized InAs QDs FET (figure 2a) is highly conductive with currents reaching above 2 mA, at a gate voltage, $Vg = +30$V, and exhibit n-type nature in-line with previous studies regarding the n-type characteristics of InAs based FETs.[22,48,31,49] The transfer characteristics of this device presented in figure 2d indicate a highly n-type 'inherent' doping of the QDs, with a threshold voltage, $V_{TH} = -23$ V and high linear mobility, $\mu_{e,lin.} = 0.53$ cm$^2$ V$^{-1}$s$^{-1}$. Bulk intrinsic InAs is known to have an electron surface accumulation layer, which originates from electron donating surface states, that effectively pin the Fermi level very close to or even above the conduction band minimum.[50] Because of the extremely high surface area to volume ratio of InAs QDs, this phenomenon is much more profound, leading to strong inherent n-type conduction.[22,31,49] Therefore, for successful p-type doping it is necessary to overcompensate for the intrinsic high concentration of electrons. FETs produced from Zn(Oleate)$_2$ – reacted QDs also shows dominant n-type characteristics, but to a lesser extent than the as-synthesized QDs, reflected from the lower conductivity in the output characteristics in figure 2b. The threshold voltage shifts to more positive values, $V_{TH} = +1.5$ V, as seen in the transfer characteristics (figure 2e) and lower linear electron mobility, $\mu_{e,lin.} = 0.003$ cm$^2$ V$^{-1}$s$^{-1}$. Notably, successful p-type doping and switching of the majority carrier type from electrons to holes was only achieved by treating the QDs with the more reactive diethylzinc as presented in figure 2c,f. In this case the current polarity is reversed, and its magnitude increases as more negative gate bias is applied,

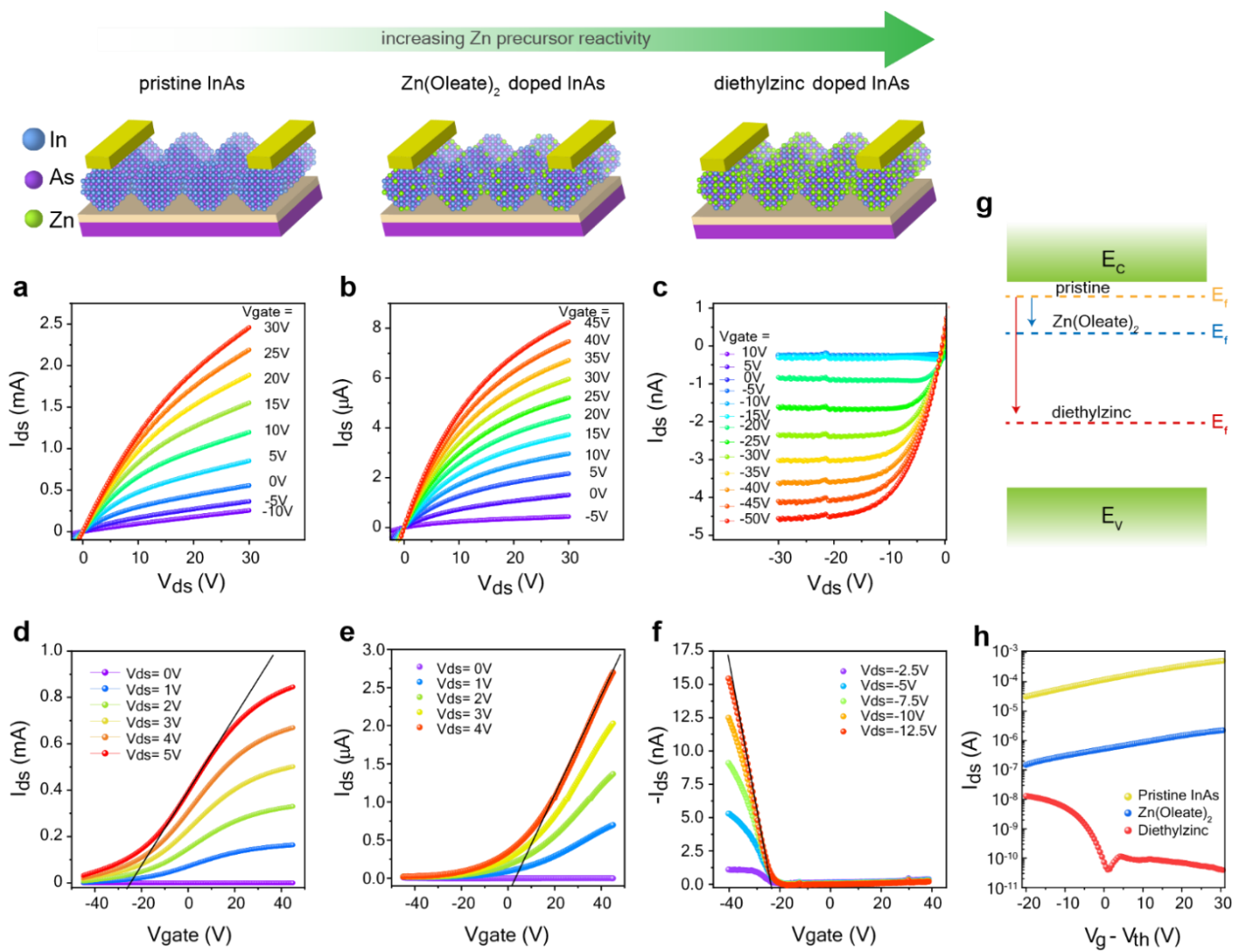

**Figure 2. Zn doping effect on InAs NCs-based devices performance**. a-c, Output characteristics of FETs composed of pristine (a), Zn(Oleate)2-doped (b), and diethylzinc-doped (c) InAs NCs exhibiting the transition from highly n-type to p-type conduction as Zn precursor reactivity is increased. d-f, Transfer characteristics of the respective devices indicating the shift in the threshold voltage from heavy n-type to low p-type conduction of the devices. g, Schematic illustration of the trend of Fermi level, $E_f$, shift relative to the conduction band, $E_c$, with respect to different Zn-dopant precursors. h, Semi-log scale plot of the transfer characteristics exhibiting the change in Ids on/off ratio with the transition from highly conductive n-type to moderate p-type doping. Note that frames a,b show the I-V characteristics in the 1st quadrant while c the 3rd quadrant, in compliance with the conventional representation and n-type and p-typed FET behavior.[22,23,51]

indicative to the conduction of holes, as presented in the output characteristics in figure 2e. The transfer characteristics of this device, seen in figure 2f, show high threshold voltage, $V_{TH} = -24$ V for opening the channel in the p-type device and much lower hole mobility, $\mu_{h,lin.} = 1.2 \times 10^{-5}$ cm$^2$ V$^{-1}$s$^{-1}$. However, the On/Off ratio improves compared to the as-synthesized and Zn(Oleate)$_2$ doped FETs, as seen in figure 2h

We can address the results from the electrical measurements of the FETs, by considering the Fermi level location relative to the conduction and valence band of the QDs film, as depicted schematically in figure 2g. The highly n-type characteristics of the as-synthesized InAs QDs film can be explained by placing the effective equilibrium fermi level very close or even pinned to the conduction band. This way, the device is in its ON state even without applying a gate voltage. By treating the InAs QDs with Zn(Oleate)$_2$, the effective fermi level is pushed away from the conduction band, due to p-type doping. However, the fermi level is still close enough to the conduction band to maintain n-type behavior, with low On/Off ratio as indicated in figure 2h. Upon doping the InAs QDs with the more reactive precursor (diethylzinc) – a substantial amount of surface dangling bonds is passivated. This, presumably, is accompanied by In-to-Zn substitution (as hypothesized by the byproducts found during the doping reaction), leading to enhanced p-type doping and pushing the effective fermi level even further away from the conduction band. Its position closer to the valence band leads to p-type nature with high (negative) threshold voltage. Overall, therefore, successful p-type doping of InAs QDs with Zn was achieved. The study clearly highlights the effect of the dopant precursor's reactivity on the successful doping of colloidal QDs. To gain further insight on the chemistry of Zn-doping and accounting for the difference in the p-type doping efficiency between the two studied precursors, we performed advanced structural and spectroscopic characterization of the doped QDs.

## 2.3. UHR-STEM and XPS Characterization of Zn-doped InAs

Figure 3a shows ultrahigh resolution STEM with energy dispersive X-ray spectroscopy (EDS) elemental mapping of Zn(Oleate)$_2$ doped InAs QDs, with very weak Zn signal on the surface of the QDs. On the other hand, diethylzinc doping induces significant changes in the elemental content of the InAs QDs. As seen from the EDS elemental mapping, in figure 3b, the coverage of Zn on the surface is much more homogenous and pronounced. To quantify the doping reaction yield, the atomic fraction of In, As and Zn in the as-synthesized and Zn doped QDs was extracted, as presented in figure 3c. The as-synthesized InAs QDs are known to have excess Indium on the surface of the QDs. We observed an In:As ratio of 1.3:1.0. This off-stoichiometry can explain the inherent n-type nature of the as-synthesized InAs QDs, as was suggested for off-stoichiometric PbSe QDs.[52] Upon doping with Zn(Oleate)$_2$, the In:As ratio decreases slightly to 1.2:1.0 with an addition of 6% Zn to the total content of the QDs, implying that doping the QDs with Zn(Oleate)$_2$ enables very limited substitution of In with Zn, not enough to induce dominant p-type doping of the QDs.

Regarding diethylzinc doping, a rise in Zn content compared to Zn(Oleate)$_2$ treatment is observed, together with a significant decrease in Indium content, becoming minority relative to As, while the In:As ratio decreases to 0.85:1.0. This is an additional strong indication of substitutional doping owing to the higher reactivity of the Zn precursor. The cross-section of the EDS image on a diethylzinc doped InAs QD (green line in bottom right of figure 3b; Figure 3d) indicates that Zinc not only resides on the surface of the QDs but can also diffuse deeper and substitute Indium. From this, we hypothesize that when the QDs are doped with Zn(Oleate)$_2$, Zinc binds to the surface of the QDs (leading to improved PL) with very little substitutional doping (hence, Zn(Oleate)$_2$ doped InAs FETs remain n-type). When the QDs are doped with diethylzinc, its higher reactivity enables Zinc to diffuse inside the QDs lattice and possibly substitute Indium atoms.

To support this hypothesis, we aimed to gain more insight on the chemistry of the doping process and characterized the as-synthesized and the doped QDs using x-ray photoelectron spectroscopy (XPS), to identify chemical changes on each of the above elements after doping. XPS is a surface sensitive characterization method and is suitable for this study, where the main chemical modifications happen at or near the surface of the QDs. The results in figure 4 indicate that upon Zn doping with either of the Zn precursors, oxides bonded to the surface of the as-synthesized QDs are removed. The existence of $In_2O_3$ on the surface of the as-synthesized QDs is revealed by the Indium 3d signal (figure 4a), where the main In-As signals at 444.4 eV and 451.9 eV are accompanied by additional In – O contributions at 445.2 eV and 452.5 eV, which are later being removed during the doping process. The same applies for the Arsenic 3d signal (figure 4b), which combines As – In (40.7 eV and 41.3 eV) and As – O (42 eV and 42.8 eV) contributions, with the oxide being removed after doping. Zn 2p signal in figure 4c of both Zn-doped samples indicates its presence but it is nearly impossible to distinguish between the different Zn oxidation states by XPS.

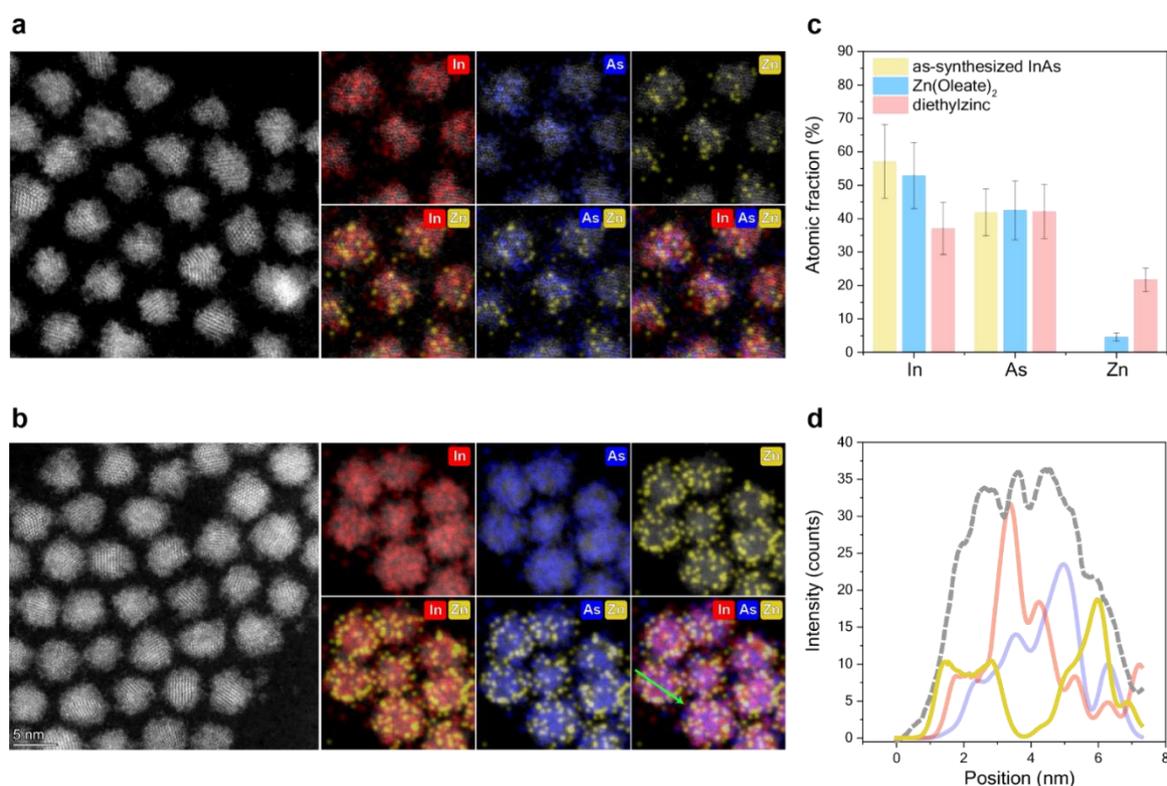

**Figure 3. Zn precursor reactivity affects elemental composition and distribution in Zn-doped InAs NCs**. a, UHR-STEM and EDS elemental mapping of Zn(Oleate)2 treated InAs, showing low quantities of surface attached Zinc, and core In, As. b, Similar data for diethylzinc treated InAs treated, showing an increased amount of Zinc. C, Quantification of the added Zinc impurities for different Zinc treatments and the decrease of Indium to Arsenic ratio as measured by EDS, assigned to Zn-to-In substitution doping. D, Cross sectional scheme for diethylzinc-doped InAs (depicted by the green line in b), implying that diethylzinc treatment enables the diffusion of Zinc from the surface into the InAs core.

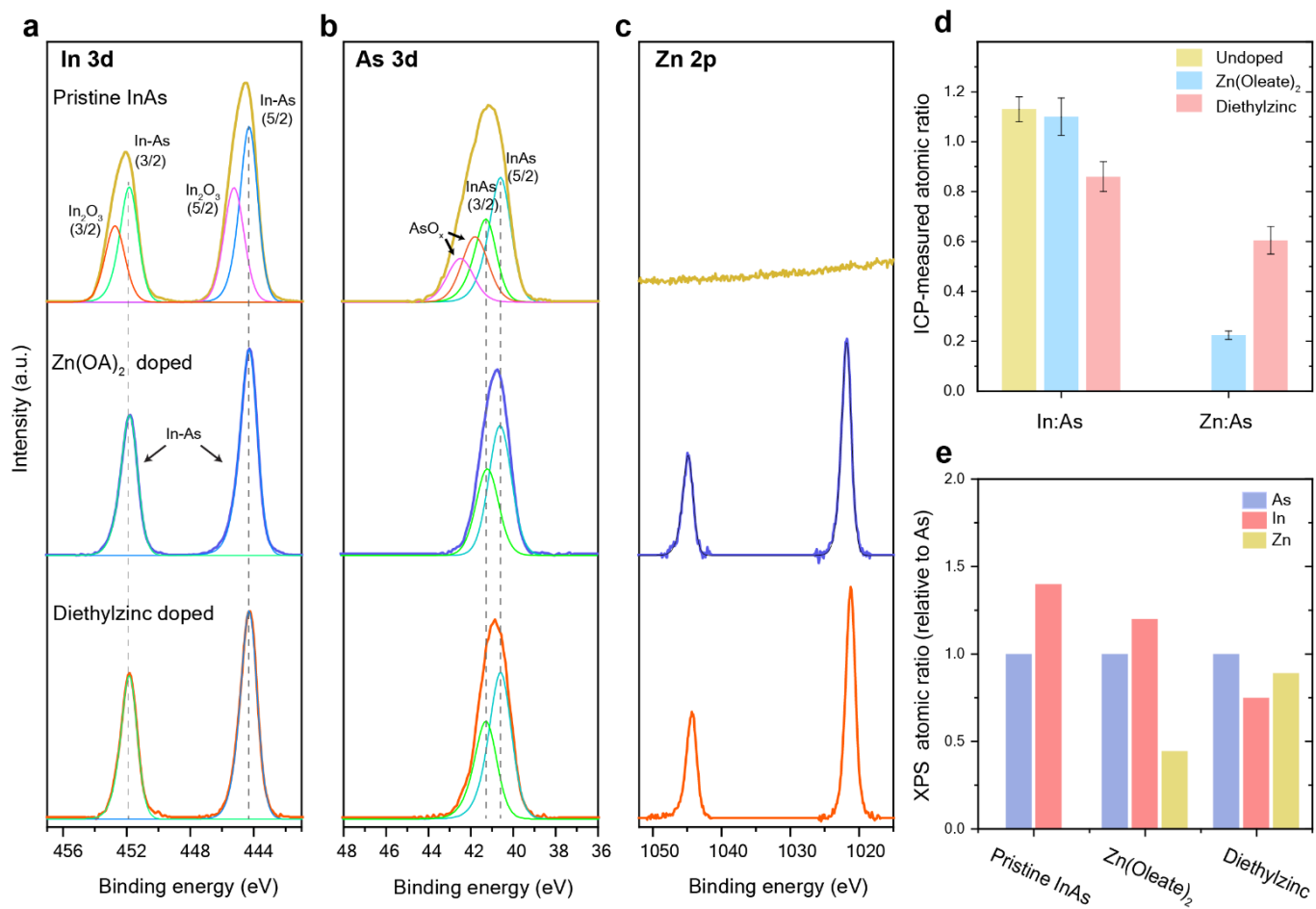

**Figure 4. Chemical effects of Zn doping in InAs NCs**. **a-c**, Xray Photoelectron Spectroscopy characterization showing In, As and Zn signals of as-synthesized (dark yellow, upper row), Zn(Oleate)$_2$ doped (blue, middle), and diethylzinc doped (red, bottom row) InAs QDs. Data is shown after baseline subtraction (raw data and baseline fits are shown in SI Figure S10). In both doping reactions, the addition of Zn removes surface InO$_x$. **d-e**, Quantification of the added Zinc impurities for different Zinc treatments and the decrease of Indium to Arsenic ratio as measured by ICP and XPS, respectively. Both results indicate that Zn substitutes In in the system with varying efficiency related to Zn precursor reactivity.

To further correlate the chemical information gained from In and As with quantitative analysis, we performed inductively-coupled plasma mass spectroscopy (ICP – MS) characterization. This allows to quantify the atomic ratios in the different systems and to compare it with the atomic fraction of each element calculated from the XPS analysis. In figure 4d, ICP – MS calculated In:As atomic ratios of 1.13:1.0, 1.08:1.0, and 0.85:1.0 for as-synthesized, Zn(Oleate)$_2$ doped-, and diethylzinc doped InAs QDs, respectively, are in line with the trend measured by EDS. The atomic ratios calculated from the XPS characterization (figure 4e) follows the same trend. However – the higher sensitivity of XPS to the surface indicates larger In:As ratios (because of In rich surface) compared to the ones acquired by ICP, which sums the overall atomic ratio (surface + core).

The removal of surface In$_2$O$_3$, as evidenced from XPS characterization, together with the decrease of In:As atomic ratio during the doping reaction, as indicated by quantitative analyses, supports our hypothesis of substitutional doping of the InAs QDs when doped with diethylzinc. However, no direct indication of substitutional doping exists from the Zinc point of view. Therefore, to confirm this hypothesis and to reveal the site-specific location of Zn within the InAs lattice, thus providing insight on the doping mechanism, we turn to atomic level structural characterization utilizing synchrotron-based XAFS measurements of the doped samples.

## 2.4. XAFS study for the analysis of the doping mechanism

XAFS measurements provide a powerful tool to study both the local and electronic structures of impurity doped colloidal nanocrystals.[20,21,23,45,53–55] Herein, XAFS K-edge spectra of In, As, and Zn were collected for as-synthesized InAs QDs, and Zn-doped QDs doped with either Zn (Oleate)$_2$ or diethylzinc, with increasing Zn concentrations. The measurements were performed at the 7-BM (QAS) beamline of the National Synchrotron Light Source – II (NSLS-II) at Brookhaven National Laboratory. For the Zn(Oleate)$_2$ doped QDs -  the dopant levels are reported by amounts equivalent to monolayers on the surface of the InAs QDs, namely: low doped (eq. to 1 ML), moderately doped (eq. to 2 ML) and highly doped InAs QDs (eq. to 3 ML). For the case of low Zn(Oleate)$_2$ doped QDs, the amount of Zn(Oleate)$_2$ added corresponds to a Zn:As ratio of 0.5:1.0. this ratio is doubled and tripled for the more concentrated doped QDs samples, respectively. For the QDs doped with diethylzinc, variances in post-synthesis doping reactions led to a wider range of doping concentrations, with Zn:As ratios after purification of the QDs ranging from 0.33:1.0 to 1.5:1.0 as measured using ICP-MS.

We first examine the In and As K-edges by comparing the effect of doping with the two different Zn precursors on the chemical environment from the host atoms' point of view. We start with the QDs doped with the more reactive precursor diethylzinc and showing p-type character in FETs. The normalized Indium K-edge XANES spectra, presented in figure S3, reveals the existence of isosbestic points, which indicate a transition between two different chemical environments surrounding In as the Zn concentration rises. We hypothesize that it can be explained by the substitution of Indium on the surface by Zinc – transitioning from a mixture of InAs and InO$_x$ phases to a single In-As interaction at the core of the QDs. This hypothesis will be tested by the Zn K-edge data analysis below. Additionally, the In K-edge white line intensity decreases upon doping, meaning that, on average, Indium is bound to a less electron-withdrawing atom, which could also be explained by the elimination of In-O bonds and the increased fraction of In-As interactions per average In atom in the doped NCs. This is corroborated by the changes seen in the Fourier transformed EXAFS signal of Indium, also presented in figure S3. As-synthesized InAs QDs show a strong peak at 2.3 Å corresponding to the interaction of Indium with Arsenic, and another sub-peak at 1.6 Å implying for In-O interaction. As the Zn concentration increases, the feature at 1.6 Å diminishes while the main peak at 2.3 Å increases. This result is further clarified when fitting the data to a model of bulk InAs with some surface Indium oxide (figure S4). In this model, depending on the location of Indium atoms, they are characterized by either a single scattering path (In-As for In in the bulk InAs, In-O for In in the surface In oxide) or two scattering paths (In-As and In-O for In in the interface location). Therefore, because EXAFS is an ensemble-averaging technique, a fitting model that included both In-As and In-O contributions was used. Upon diethylzinc doping, the In-O path is indeed eliminated from the EXAFS, shifting the weights toward a purer InAs phase from the Indium perspective. To get a clearer view of the change in the Indium surroundings with different doping procedures, the coordination number gained from the model fit with respect to Zn doping levels is displayed in figure S5. First, it is seen that for as-synthesized InAs QDs, the coordination number (C.N.) of In with surrounding As is lower than the value expected for a stoichiometric InAs, in agreement with In-rich QDs that have surface Indium atoms that are under coordinated or oxidized.

For InAs doped with diethylzinc, the coordination number increases and gets closer to the stoichiometric value of 4 As neighbors in all concentrations, along with the elimination of the coordination number of In with surrounding Oxygen. This shows that surface Indium is being removed after the doping reaction. In contrast to Indium, Arsenic K-edge EXAFS spectra show no changes upon doping, as indicated in figure S6.

Less significant changes in In and As K-edge XAFS data are evidenced when the $Zn(Oleate)_2$ precursor was used for the doped samples (figure S7). These observations imply that no significant change occurs in the surrounding chemical environment, regardless of doping concentration. The Fourier transform magnitude of In $k^2$-weighted EXAFS data and of as-synthesized and $Zn(Oleate)_2$-doped InAs QDs samples (Figure S8) and fit to the same model described above shows that no significant changes in the chemical environment are seen in Indium upon doping with $Zn(Oleate)_2$, except for a slightly smaller contribution of In-O interaction to the total fit, which supports the existence of surface Indium oxide even after the doping procedure. This could be a result of the low reactivity of $Zn(Oleate)_2$, causing the removal and substitution of surface In with Zn to be inefficient. In contrast to the diethylzinc precursor, doping the QDs with Zn(Oleate)$_2$ results in a gradual increase in In-As CN and a gradual decrease in In-O C.N.

Regarding As K-edge, in the XANES region (figure S7), only a minor increase in white line intensity was observed after doping. The Fourier transformed EXAFS signals of all samples are analyzed by a model that includes only a single As-In scattering path (figure S9), indicating that the chemical environment surrounding Arsenic remains unchanged regardless of Zn doping concentration. To further decipher the doping mechanism and differences between the two different precursors, we turn the attention to the dopant perspective, presented in figure 5.

The Zn K-edge data provide key information on the nature of doping and the difference between the performance of the two precursors. Zn K-edge XANES spectra of diethylzinc doped samples presented in figure 5a shows strong sensitivity to doping, and significant changes in the chemical environment surrounding Zn appear as the Zn content increases. It is evident that XANES of the low concentration Zn-doped samples strongly resembles that of bulk ZnO.[56] The main dominant peak decreases while an additional feature develops at lower energy as the Zn content increases. This means that Zn is no longer predominantly in the form of a ZnO phase, and a more complex chemical environment exists when the dopant concentration increases. EXAFS spectra further emphasize the effect of diethylzinc doping (Figure 5c) on the local environment of Zn. All doped samples show a strong peak at a radial distance of ~1.5 Å and additional peak at ~3 Å. These peaks are related to the backscattering of the photoelectrons from the nearest neighbors of the first and second coordination shell of Zn and are assigned to Zn-O and Zn-Zn interactions of ZnO. However, as diethylzinc doping concentration increases, an additional feature develops at 2 Å, which is likely to be associated with a Zn-As interaction due to substitutional In-to-Zn doping. Importantly, samples that share

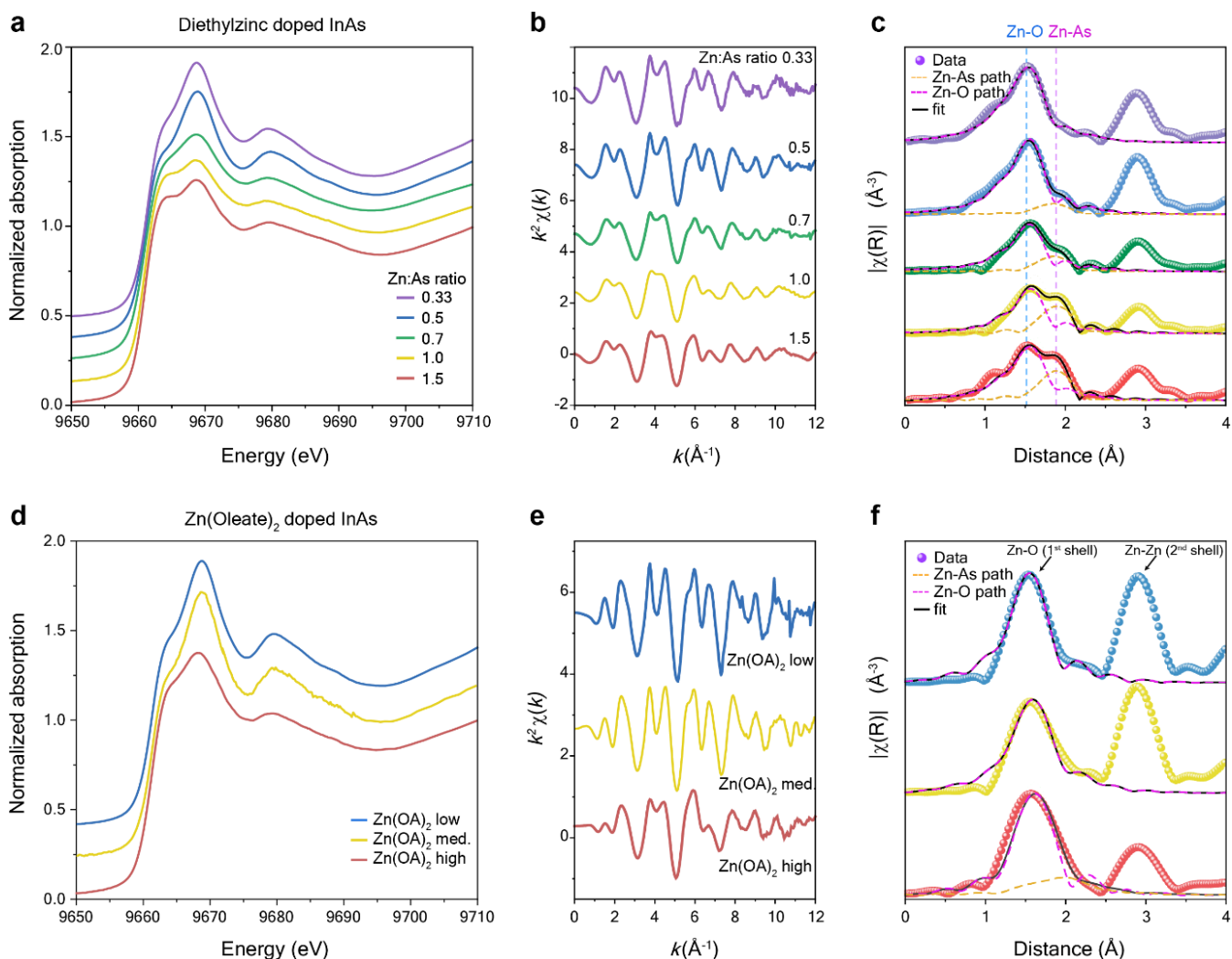

**Fig. 5 | Precursor reactivity effect on the local environment and bonding of Zn within InAs NCs lattice**. **a**, Normalized Zn K-edge XANES spectra of InAs NCs treated with increasing concentrations of Zn(Oleate)$_2$ exhibiting ZnO phase. **b**, $k^2$-weighted Zn K-edge EXAFS spectra of the respective samples. **c**, Fourier transform magnitude of Zn K-edge EXAFS data (spheres) of Zn(Oleate)$_2$ doped InAs samples (k-range: 2-13 Å$^{-1}$; rbkg: 1.23) plotted together with a fit of the first coordination shell (black line) using a structural model of a Zn impurity forming an oxide layer on the surface of InAs NCs. Zn-O and Zn-As scattering paths are depicted as purple and dark yellow dashed lines, respectively. **d**, Normalized Zn K-edge XANES spectra of InAs NCs treated with increasing concentrations of diethylzinc exhibiting a transformation from a pure ZnO phase to a mixture of ZnO and Zn-As. **e**, $k^2$-weighted Zn K-edge EXAFS spectra of the respective samples treated with different concentrations of diethylzinc. **f**, Fourier transform magnitude of Zn K-edge EXAFS data (spheres) of diethylzinc doped InAs NCs (k-range: 2-13

Å$^{-1}$; rbkg: 1.23) plotted together with a fit of the first coordination shell (black line) using a structural model of a Zn impurity occupying surface site (Zn-O path, purple dashed line), together with an additional substitutional site as diethylzinc concentration increases (Zn-As path, yellow dashed lines). this additional peak also contain the highest amount of Zn, according to ICP-MS. In contrast, the Zn K-edge XANES and EXAFS data for the samples prepared using Zn(oleate)$_2$ precursor paint a completely different picture. Figure 5d shows the Zn K-edge XANES spectra of the Zn(Oleate)$_2$-doped InAs QDs with increasing doping concentrations. It is evident that XANES of the low and medium Zn-doped samples strongly resembles that of bulk ZnO with only that of the highly Zn(Oleate)$_2$ doped sample slightly changing, indicating only small changes in the chemical environment surrounding Zn. In the Fourier transformed EXAFS spectra, Zn(Oleate)$_2$ doped samples show the main Zn-O and Zn-Zn peaks from the ZnO structure at ~1.5 Å and ~3 Å, respectively. This suggests that in this case Zn is simply bound to the surface of InAs and is dominantly in the form of ZnO.

We support our qualitative interpretation of both doping models by quantitative fitting analyses of the EXAFS data. As informed by our visual examination of the raw data, for the sample doped using diethylzinc precursor, we used a model describing a combination of Zn as a substitutional dopant for In in the bulk of the QDs and of surface Zn forming ZnO, described by contributions from both Zn-O and Zn-As scattering paths. We hypothesize that given that the as-synthesized InAs QDs surface is Indium rich (as indicated by ICP-MS), Zn substitutes In atoms in the bulk of the QDs and will thus interact with the nearest neighboring As atoms. In this scenario, remaining Zn cations that are exposed to the surface of the QDs and will oxidize in the presents of small amounts of oxygen and the high temperature during the reaction.

It is evidenced that for samples that contain a low concentration of Zn, using only the Zn-O path is sufficient to fit the data presented in figure 5c. However, as the Zn content within the samples increases, the Zn-As path must be considered as well to fit the data, and this path becomes increasingly more significant for higher Zn concentrations. For the Zn(oleate)$_2$ precursor-doped samples, to fit the first coordination shell of the EXAFS data, we used a single Zn–O scattering path only (figure 5f). The model fits the low and moderate dope samples data well. For the highly doped sample, an additional Zn-As scattering path was included, however, its contribution did not affect the overall fit significantly, compared to the fit of the diethylzinc precursor-doped material. The overall picture emerges – for the diethylzinc doped samples, the higher the Zn concentration is, the more substitutional doping occurs, as depicted in figure 6a, plotting the Zn-O and Zn-As coordination numbers acquired from the fit results and illustrated in the inset. For the Zn(Oleate)$_2$ doped InAs samples, the existence of Zn-As interaction is only evidenced in the highly doped sample with an only minor contribution, as presented in figure 6b.

The above discussed multi-technique analysis demonstrated that, upon the doping of InAs QDs with diethylzinc, Zn diffuses into the core of the QDs, leading to substitutional doping and p-type conduction in FETs. In contrast, in Zn(Oleate)$_2$ – doped QDs, Zn predominantly attaches to the surface of the QDs while also replacing some of the In on the InO$_x$ phase. Zn also resides on the surface upon the diethylzinc treatment as observed from STEM, XPS and XAFS results. This already significantly lowers the "doping efficiency", namely the extent of free carriers actually induced relative to the Zn amount. The surface Zn is not contributing free carriers. Additionally, the original high n-type character of the pristine NCs must be overcompensated to turn over to p-type conduction. This limits the doping efficiency as well. There may also be additional factors in such small NCs, which decrease doping efficiency. This may include formation of defects, leading even to band tailing in the heavy doping limit.[19] The doping efficiency as expressed in the FET channels was also very low also in Cu doped n-type

InAs NCs,[22] suggesting that this is a more general characteristic in nanocrystals. This calls for further studies of this aspect.

## 3. Conclusion

In summary, we developed a post-synthesis doping reaction to achieve heavy metal-free p-type InAs QDs. Using a combination of electron microscopy, XPS, and XAFS characterizations, the location of Zn within the InAs lattice was identified, gaining an understanding on the Zn-precursor's reactivity-dependent doping mechanisms. This directly accounts for the variations in the electrical performance of the Zn-doped FETs. Control of the doping process in III-V colloidal QDs enables a strategy for bottom-up fabrication of QDs-based p-n homojunction opening the path for a multitude of all-nanocrystal-based electrical devices. This study may thus serve as a basis for future development of colloidal III-V QDs-based optoelectronics.

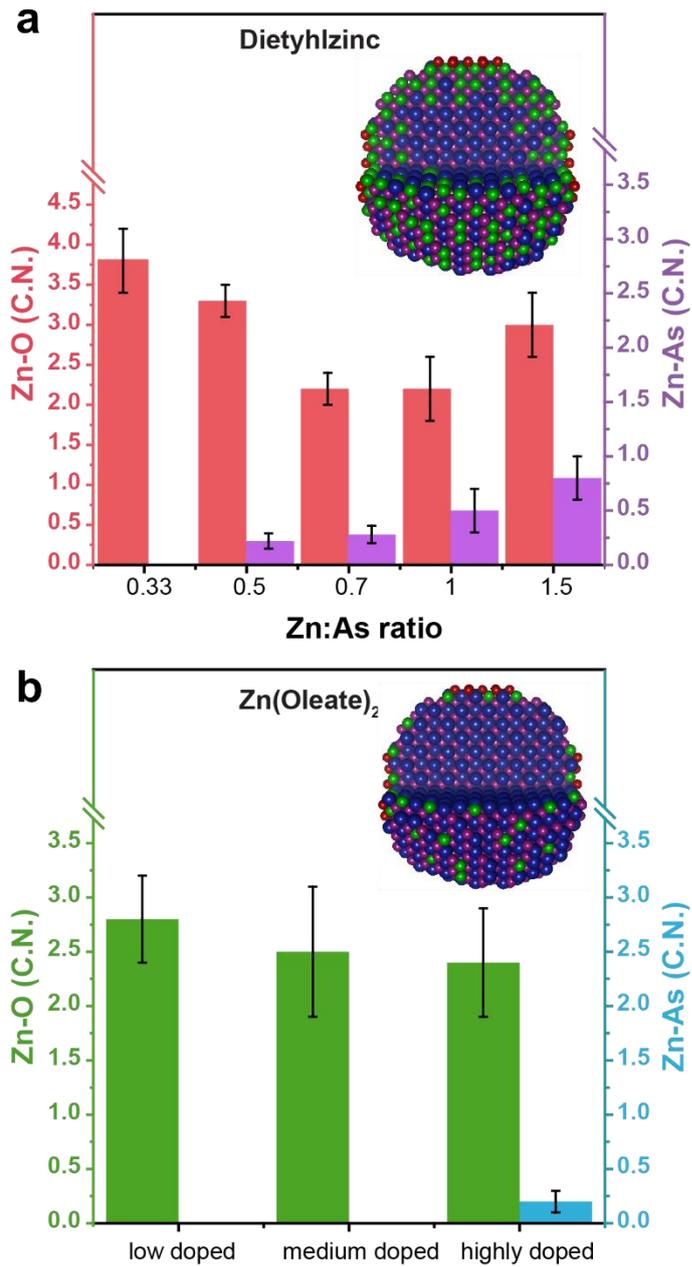

**Figure 6.** The coordination number of Zn with surrounding O and As, respectively, for InAs samples doped with diethylzinc (**a**), and Zn(Oleate)$_2$ (**b**). The insets in both panels represent a structural illustration to describe the location of Zn within the QDs lattice.

## 4. Experimental

*Synthesis of Colloidal InAs NCs:* Colloidal InAs NCs were synthesized following a well-established continuous injection method: InAs nanocluster solution was prepared by mixing 12 mmol of In(OAc)$_3$ and 36 mmol of Oleic acid with 60 mL of ODE in a 100 mL 3 neck flask. The mixture is degassed at 120c under vacuum for 90 minutes and than left to cool down to room temperature under Ar. In the glovebox, 1.6 g of distilled (TMSi)$_3$As is mixed with 12 mL of dry ODE. The As solution is injected at room temperature into the Indium-oleate solution with constant stirring. The temperature is slowly raised to 50C, until the solution color turns dark red. The InAs nanocluster is cooled down to room temperature and is kept under inert atmosphere in the glovebox. To a new 5o mL 4 neck flask, 2 mmol of In(OAc)$_3$ and 6 mmol of Oleic acid are mixed together with 5 mL of ODE and left under vacuum at 120C for 2 hrs, and then switched to Ar and the temperature is raised to 300C. In the glovebox, 0.24 g of distilled (TMSi)$_3$As is mixed with 2 mL of dry ODE. The As solution is injected at 300C into the Indium-oleate solution with constant stirring to form InAs seeds. The temperature is lowered to 285C and remain constant for about 20 minutes. During this time, aliquots are taken and their absorption spectrum is measured. Once the 1$^{st}$ exciton absorption feature of the seeds reaches ~ 750 nm, the nanoclusters solution is added continuously at a constant rate of 35 $\mu L/min$. During this reaction aliqouts are taken at constant intervals to evaluate the growth o size focusing of the nanocrystals by absorption measurements. At the end of the synthesis, the cooled solution is transferred into the glovebox and successive precipitation – centrifugation – redispersion cycles are performed in order to clean the nanocrystals from the crude solution using dry hexane and butanol as solvent and anti-solvent, respectively. The clean nanocrystals are dispersed in hexane and kept in the glovebox for further use.

*Post-synthesis doping of InAs QDs*: mL of ODE and 1.5 mL of OlAm were added to a 50 mL round bottom flask connected to Schlenk line and were heated under vacuum to 100°C for 1 hr. 100 nmol of highly monodispersed fraction of as synthesized InAs QDs in hexane was added to the flask and the solution was evacuated at room temperature for 30 minutes and then the temperature was raised again to 70 °C, still under vacuum for additional 30 minutes in order to remove any of the hexane residues. Then, the temperature was raised to 260°C under Ar flow and once the temperature settled, calculated amount of Zn(Oleate)$_2$ in ODE (0.2M) or diethylzinc (0.1 M in degassed ODE), correlating to overall In:Zn ratio of 1:1.2 were added dropwise over a period of 20 minutes. The system was left for 3 hrs. during this time, aliquots were taken and the absorbance and photoluminescence of the doped NCs were monitored. Purification of the doped QDs from the crude solution was performed in the glovebox. The crude solution, was diluted with 5 mL of hexane and then centrifuged at 3500 rpm for 5 minutes. The precipitate containing byproducts was discarded and the clean dispersion transferred back to the glovebox. There, 5 mL of dry ethanol were added, and the solution was centrifuged at 3500 rpm for 5 minutes, the precipitate was kept and redispersed in 4 mL of hexane. This process was repeated twice more.

*FET substrates cleaning:* Heavily doped p-type Si substrates covered with 100 nm thick SiO$_2$ and 10 nm HfO$_2$ coating were cleaned by subsequently dipping and sonicating the substrate in acetone, methanol and isopropanol for 5 min. After this cleaning procedure, the substrates were blown dry with nitrogen and put inside a UV-Ozone cleaner for 30 min. The cleaned substrate were then transferred into the N$_2$ glovebox, where they were soaked in a solution of 5 mM (3-mercaptopropyl)trimetoxysilane in methanol, and kept in this solution overnight.

*CQD Layer Deposition and FET Fabrication:* Thin films of InAs CQDs on top of the above substrates were prepared as follows: 25 mg/mL solution of InAs QDs in hexane was prepared and filtered twice using a 0.1 µm diameter porous PTFE filter to remove large aggregates and impurities. Clean InAs QDs solution was added dropwise (~30 µL) on the Si/SiO2 substrate and spin-coated at 2000 RPM for 30 s. Solid-state ligand exchange with 1,2-Ethanedithiol (EDT) was performed by covering the film with 5 mM solution of EDT in acetonitrile for 30 s and spinning at 2000 rpm for 15 s to dry the film. The film was washed twice with pure acetonitrile and spinning the substrate at 2000 rpm for another 30 s. The entire cycle was repeated until the desired thickness (30−35 nm) of InAs NCs was achieved. Thermal annealing of the as prepared films was performed prior to metal contacts evaporation. As-prepared films of InAs NCs were put inside an oven inside the glovebox and heated to 220 °C for 30 min under a $N_2$ environment. During this process, organic residues are removed from the film and the film hardens while avoiding sintering of the QDs. Au source/drain electrodes (120 nm thick, interdigitated configuration; W = 20 mm, L = 100 μm) were thermally evaporated using Electron Beam evaporation of Au through a specially designed evaporation mask at a rate of 1 Å/s.

*FET Characterization:* Electrical characterization of the devices was performed under vacuum using a sealed probe station, in the dark. Measurements were taken using two Keithley 2400 source meters for source – gate, and source – drain bias, respectively.

Details on the optical and spectroscopic characterizations are provided in the Supporting Information.

## Supporting Information

Supporting Information is available from the Wiley Online Library or from the author.

## Acknowledgements

The research leading to these results has received financial support from the European Research Council (ERC) under the European Union's Horizon 2020 research and innovation program (grant agreement No [741767], advanced investigator grant CoupledNC; U.B.). This material is based upon work supported by the US National Science Foundation under Award 2102299 to A.I.F. This research used beamline 7-BM (QAS) of the National Synchrotron Light Source II, a US DOE Office of Science User Facility operated for the DOE Office of Science by Brookhaven National Laboratory under Contract No. DE-SC0012704. Beamline operations were supported in part by the Synchrotron Catalysis Consortium (US DOE, Office of Basic Energy Sciences, Grant No. DE-SC0012335). We thank Drs. Lu Ma, Steven Ehrlich and Nebojsa Marinkovic for their help with the beamline measurements at the QAS beamline. We thank Dr. Vitaly Gutkin and Dr. Sergei Remennik from the Unit for Nanocharacterization of the Center for Nanoscience and Nanotechnology at the Hebrew University of Jerusalem for assistance in the materials characterization and for helpful discussions. We also acknowledge the technical support of the staff of the Unit for Nanofabrication at the Center for Nanoscience and Nanotechnology at the Hebrew University of Jerusalem. U.B. thanks the Alfred & Erica Larisch memorial chair.

## Conflict of Interest

The authors declare no conflict of interest.

# Supporting Information

# Zn-doped P-type InAs Nanocrystal Quantum Dots

*Lior Asor,*[1,2] *Jing Liu,*[3] *Shuting Xiang,*[4] *Nir Tessler,*[5] *Anatoly I. Frenkel,*[4,6] *Uri Banin*[1,2]

[1] The Institute of Chemistry and [2] The Center for Nanoscience and Nanotechnology,

The Hebrew University of Jerusalem, Jerusalem 91904, Israel

[3] Physics Department, Manhattan College, Riverdale, New York 10471, USA

[4] Department of Materials Science and Chemical Engineering, Stony Brook University, Stony Brook, New York 11794, USA

[5] The Zisapel Nano-Electronics Center, Department of Electrical Engineering,

Technion – Israel Institute of Technology, Haifa 32000, Israel

[6] Chemistry Division, Brookhaven National Laboratory, Upton, New York 11973, USA

## Materials

Toluene (anhydrous, 99.8%), hexane (anhydrous, 99.8%), Methanol (anhydrous, 99.8%), Butanol (anhydrous, 98%), acetonitrile (anhydrous, 99.8%), (3-mercaptopropyl)trimethoxysilane (95%), 1,2-Ethanedithiol (98%), Indium(III) Acetate (In(OAc)$_3$, 99.999%), Zinc Acetate (Zn(OAc)$_2$, 99.99%), Diethylzinc (1M in Hexane), Oleic acid (90%), 1-octadecene (ODE, 90%), Oleylamine (OlAm, 98%) were purchased from Sigma-Aldrich (Merck) and used as received. Tris(trimethylsylil)arsine ((TMS)$_3$As) was synthesized and distilled in our lab by using a well-established procedure.

## Characterization

Absorption spectra were measured using a Jasco V-570 UV-Vis-NIR spectrophotometer. Fluorescence spectra and ensemble lifetimes were measured with a fluorescence spectrophotometer (Edinburgh instruments, FL920).

Transmission electron microscopy (TEM) was performed using a Tecnai G2 Spirit Twin T12 microscope (Thermo Fisher Scientific) operated at 120 kV. High resolution TEM (HRTEM) measurements were done using Tecnai F20 G2 microscope (Thermo Fisher Scientific) with an accelerating voltage of 200 kV.

High-resolution scanning-transmission electron microscopy (STEM) imaging and elemental mapping was done with Themis Z aberration-corrected STEM (Thermo Fisher Scientific) operated at 300 kV and equipped with high angular annular dark field detector (HAADF) for STEM and Super-X energy dispersive X-Ray spectroscopy (EDS) detector for high collection efficiency elemental analysis. The images and EDS maps were obtained and analyzed with Velox software (Thermo Fisher Scientific).

ICP−AES measurements were carried out using a PerkinElmer Optima 3000 to determine dopant concentrations within the NCs. Samples were prepared by dissolving the doped NCs in HNO3 and diluting the solution with triply distilled water.

**XAFS Measurements** were performed at the Brookhaven National Laboratory National Synchrotron Light Source II beamline 7-BM. The specimens were prepared under inert conditions in a $N_2$ glovebox, where they were spread on Kapton film, sealed and mounted on a sample holder. The holder was put inside a zip bag, filled with nitrogen to maintain inert conditions through the experiments. Experiments at the As and In K-edges were performed in transmission mode, and for the Zn K-edge, both in transmission and fluorescence. 30 consecutive measurements were taken at each edge to improve the signal-to-noise ratio in the data. XAFS data analysis and processing were performed using standard techniques. Briefly, the data were aligned in energy using a reference spectrum collected in standard materials simultaneously with the samples data, and then averaged. Background function subtraction and normalization were done using the Athena program.[1] All fittings to the acquired data were done using Artemis program.[1,2] For improved reliability of the fit - for each absorption edge, the fitting process was performed concurrently for different samples of the same edge. For Zn K-edge EXAFS fits, Zn-As, and Zn-O paths were included in the fitting model for the Zn-doped InAs QDs powders. The reduction factor ($S_0^2 = 0.839$) was obtained by fitting Zn foil and fixed for fitting all the Zn-InAs samples. Variables in the theoretical EXAFS signal included the coordination numbers of closest neighbors of a certain type around the absorbing Zn atom (N), the bond distance between the absorber and the nearest neighbors (R), and the mean-square-displacement ($\sigma^2$). $S_0^2$ was fixed for all samples while the other parameters varied. For In K-edge EXAFS fits, two scattering paths were used for fitting (In-As and In-O). For As K-edge EXAFS fits, only one path was used for fitting (As-In). both In and As K-edges coordination number was constrained to be 4 for the undoped sample in order to acquire the $S_0^2$ factors (0.66 for As and 0.82 for In). Similarly to Zn edge data, for the doped samples, N, R and $\sigma^2$ were set as variables, while $S_0^2$ was fixed for all samples. The fitting range in k space is 2-13 Å$^{-1}$ for Zn, 2-14 Å$^{-1}$ for In and 2-12 Å$^{-1}$ for As.

**Mobility calculation**

The FET mobility values were calculated using the following equation,

$$\mu = \frac{\left[\frac{dIds}{dVg}\right]}{\frac{W}{L}CiVds},$$

where W is the channel width = 2 cm, L is the channel length = 0.01 cm, and Ci (3.41 x10-4 F/m2 ) is the combined capacitance between the channel and the back gate per unit area . Vds is the drain source voltage from which the slope (dIds/dVg) was calculated.

**Carrier density calculation**

Carrier density of the NCs FETs was estimated by extracting the conductivity per unit area, σ, from the slope of the linear regime of Ids vs. Vds at Vgs = 0V. by combining this value with the as calculated carrier mobility, the carrier density can be calculated using the formula:

$$n = e * \mu * \sigma$$

Where $e$ is the elementary charge $\pm 1.6 * 10^{-19}\ C$.

The carrier densities of the undoped and Zn- doped FETs were estimated as $5.2 \times 10^{16}\ electrons/cm^3\ and\ 1.2 \times 10^{16}\ electrons/cm^3$ for the intrinsic and Zn(Oleate)$_2$-doped FETs, respectively, and $1.3 \times 10^{15}\ holes/cm^3$ for the p-type, diethylzinc doped FET.

# Figures and Tables

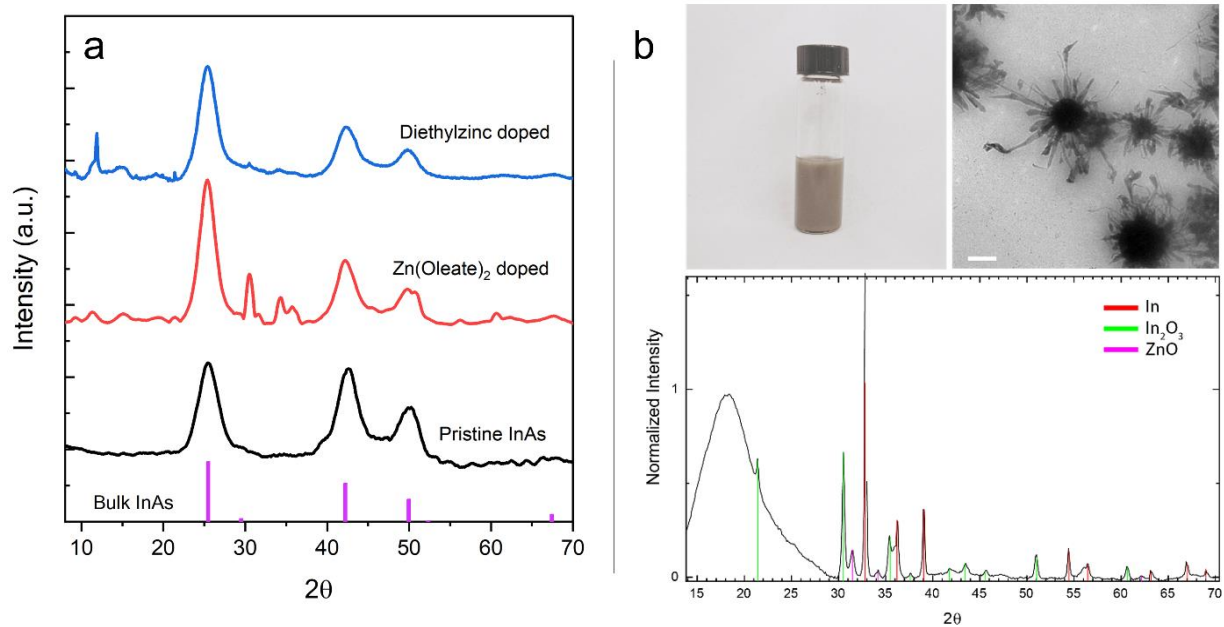

Figure S1. XRD characterization of the as-synthesized and purified Zn-doped InAs QDs samples (a). TEM image and XRD characterization of the byproduct's dispersion, that was separated out of the crude solution of the Zn-doped InAs QDs (b). The scale bar in the TEM image is 200 nm.

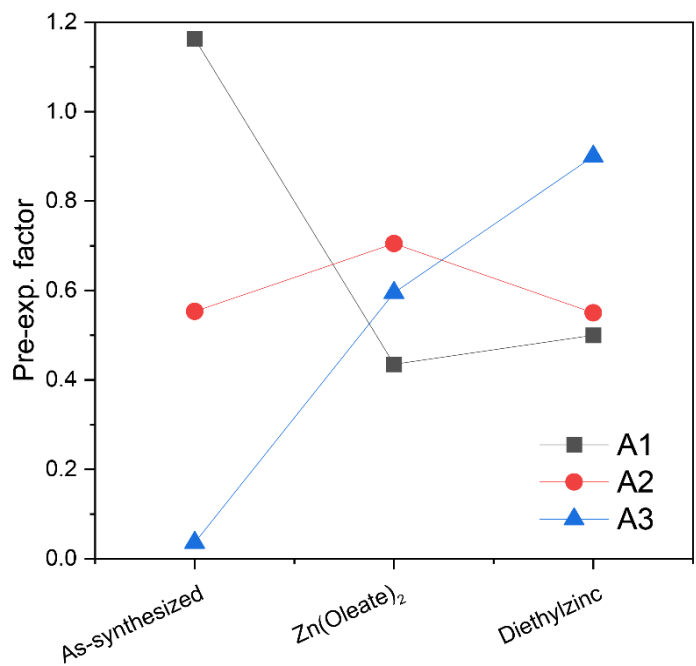

Figure S2. The change in the lifetime coefficients of the time resolved PL decay upon doping the as synthesized InAs NCs with Zn. $A_1, A_2, A_3$ correspond to the pre-exponential factors of the slow, moderate, and fast lifetime decay constants, respectively.

**Table S1. List of all the samples measured in XAFS**

| # | Sample Label | Composition | element/edge 27940 eV | element/edge 11867 eV | element/edge 9659 eV |
|---|---|---|---|---|---|
| 1 | InAs | undoped InAs NCs | In/In-K | As/As-K | |
| 3 | DEZn X3 A | | In/In-K | As/As-K | Zn/Zn-K |
| 4 | DEZn X3 B | InAs doped with high concentrations of DiethylZinc | In/In-K | As/As-K | Zn/Zn-K |
| 5 | DEZn X3 C | | In/In-K | As/As-K | Zn/Zn-K |
| 6 | DEZn X2 | InAs doped with moderate concentration of DiethylZinc | In/In-K | As/As-K | Zn/Zn-K |
| 7 | DEZn X1 A | InAs doped with low concentration of DiethylZinc | In/In-K | As/As-K | Zn/Zn-K |
| 8 | DEZn X1 B | | In/In-K | As/As-K | Zn/Zn-K |
| 11 | Zn(Oleate)$_2$ X3 | InAs doped with high concentration of Zn(Oleate)$_2$ | In/In-K | As/As-K | Zn/Zn-K |
| 12 | Zn(Oleate)$_2$ X2 | InAs doped with moderate concentration of Zn(Oleate)$_2$ | In/In-K | As/As-K | Zn/Zn-K |
| 13 | Zn(Oleate)$_2$ X1 | InAs doped with low concentration of Zn(Oleate)$_2$ | In/In-K | As/As-K | Zn/Zn-K |

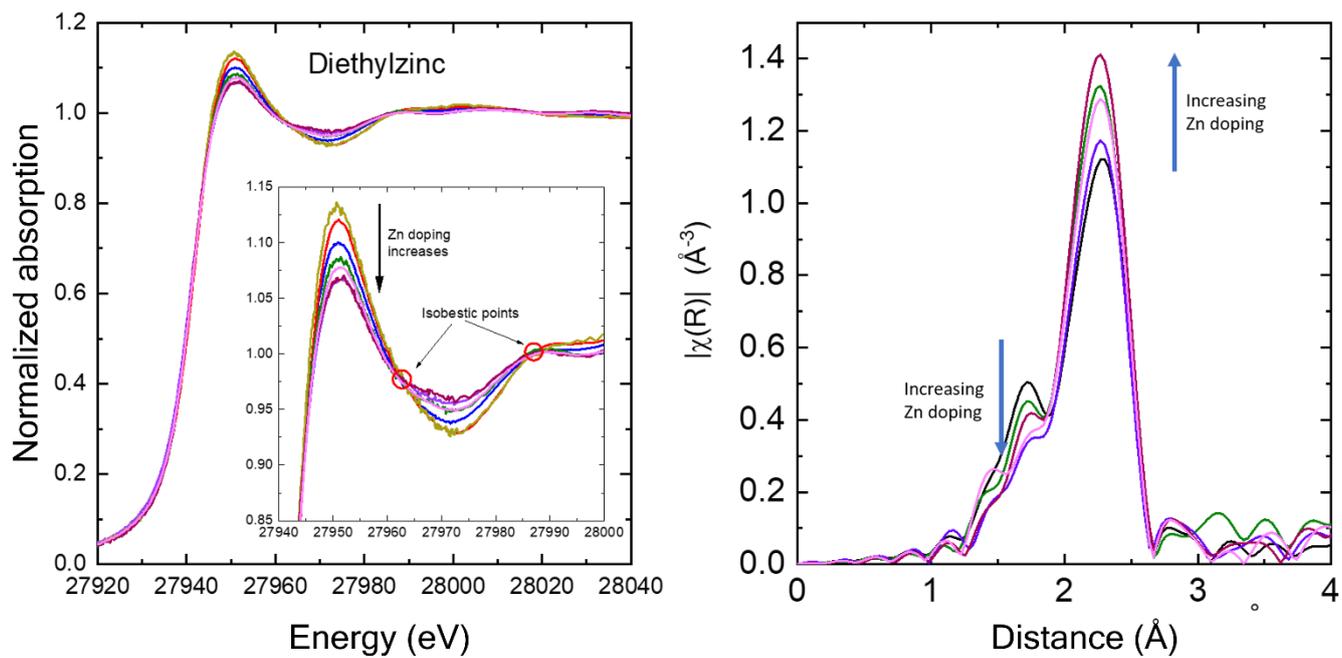

Figure S3. In K-edge normalized XANES spectra and Fourier transform magnitude of In $k^2$-weighted EXAFS data of Zn(Oleate)$_2$ doped InAs samples

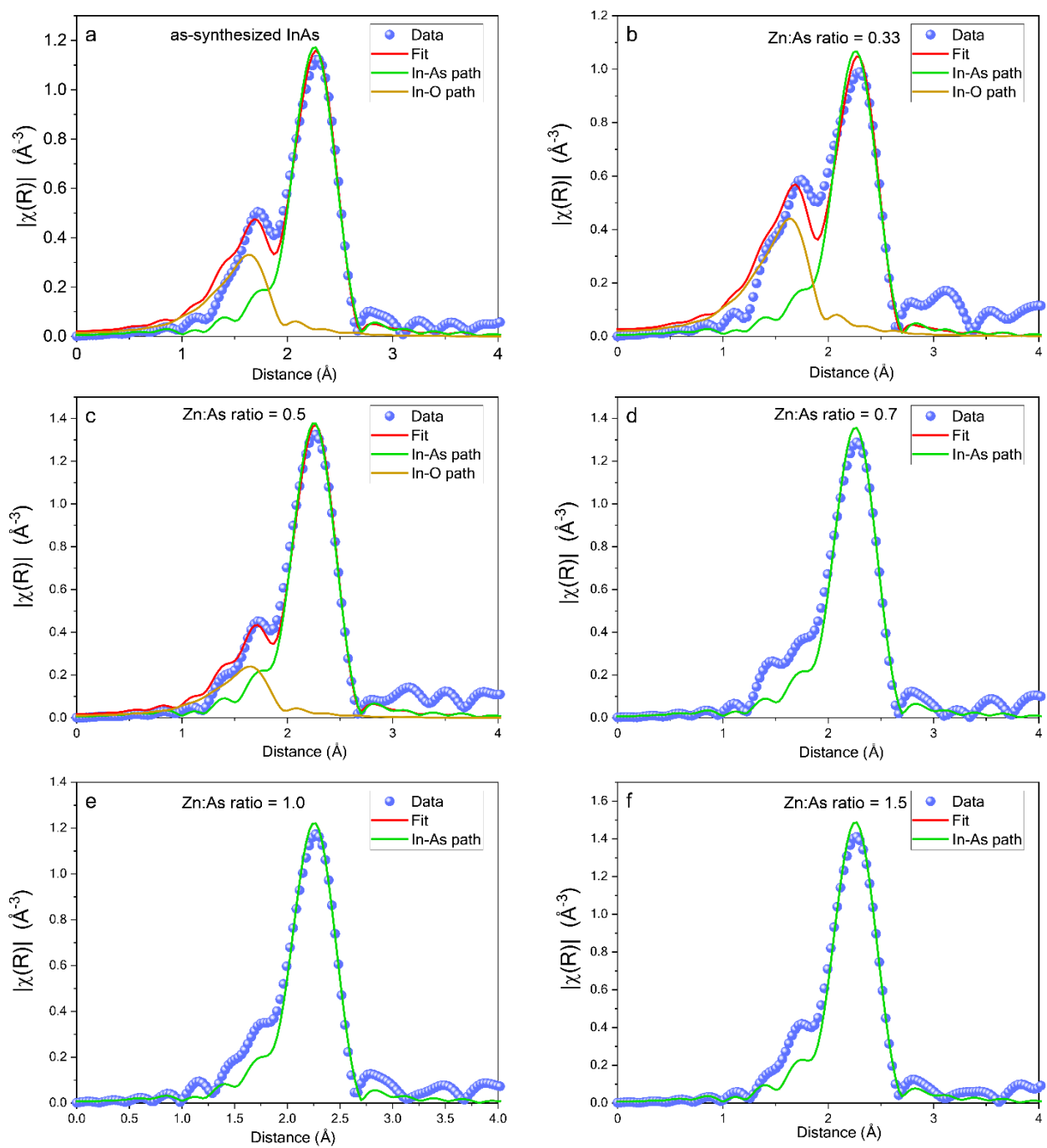

Figure S4. Fourier transform magnitude of In $k^2$-weighted EXAFS data of as-synthesized and Zn-doped InAs QDs samples doped with Diethyl Zinc (k-range: 2–14 Å$^{-1}$; $r_{bkg}$: 1.3 Å). Data (blue spheres) and fit (red line) to a structural model of Indium with two scattering paths of In-O and In-As (dark yellow and green lines, respectively)

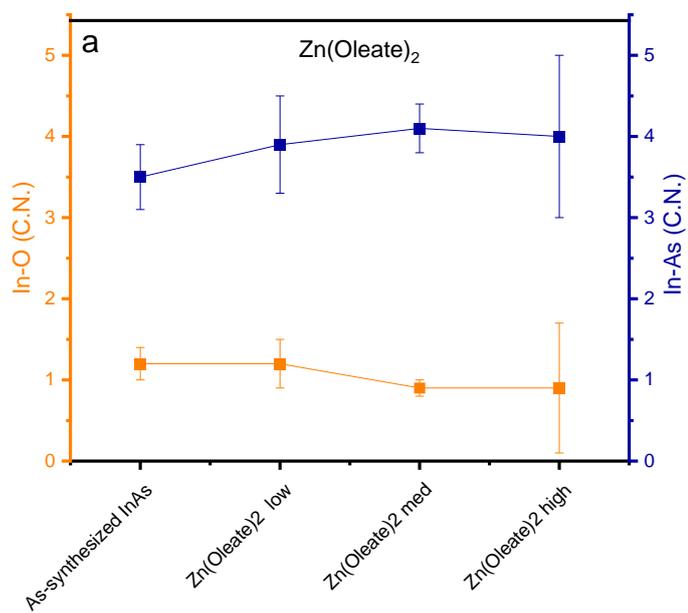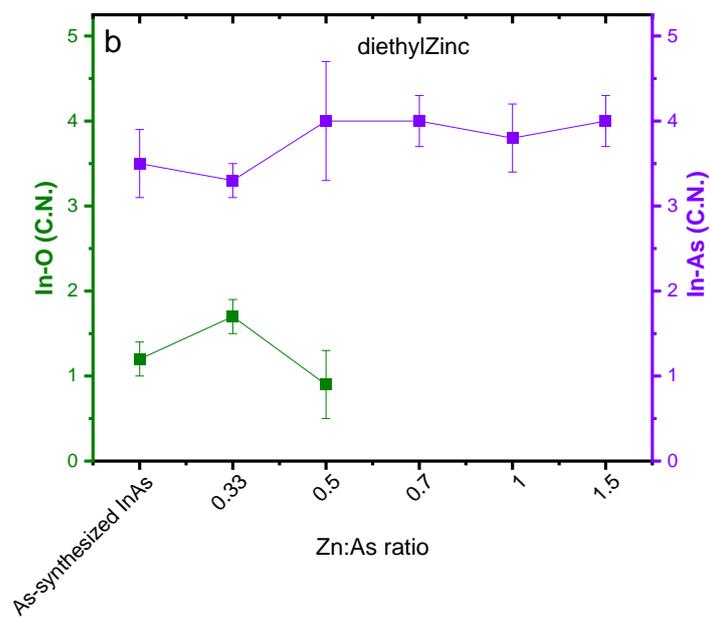

Figure S5. The fitted coordination number of In with surrounding O and As, respectively for InAs samples doped with Zn(Oleate)$_2$ (a), and Diethyl Zinc (b)

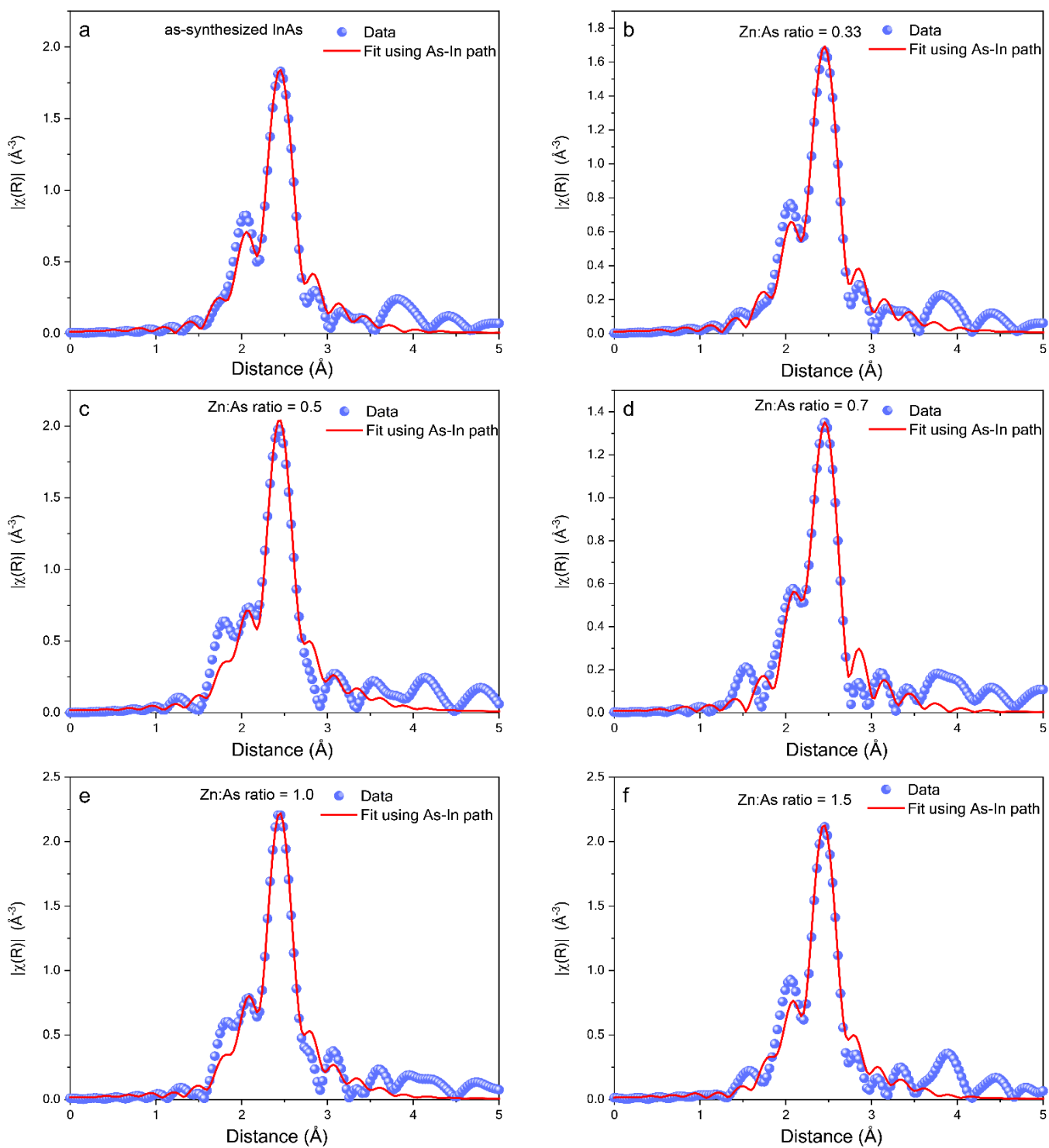

Figure S6. Fourier transform magnitude of As $k^2$-weighted EXAFS data of as-synthesized and Zn-doped InAs QDs samples doped with Diethyl Zinc (k-range: 2–12 Å$^{-1}$; rbkg: 1.7). Data (blue spheres) and fit (red line) to a structural model of Indium with a single scattering path of As-In (green lines)

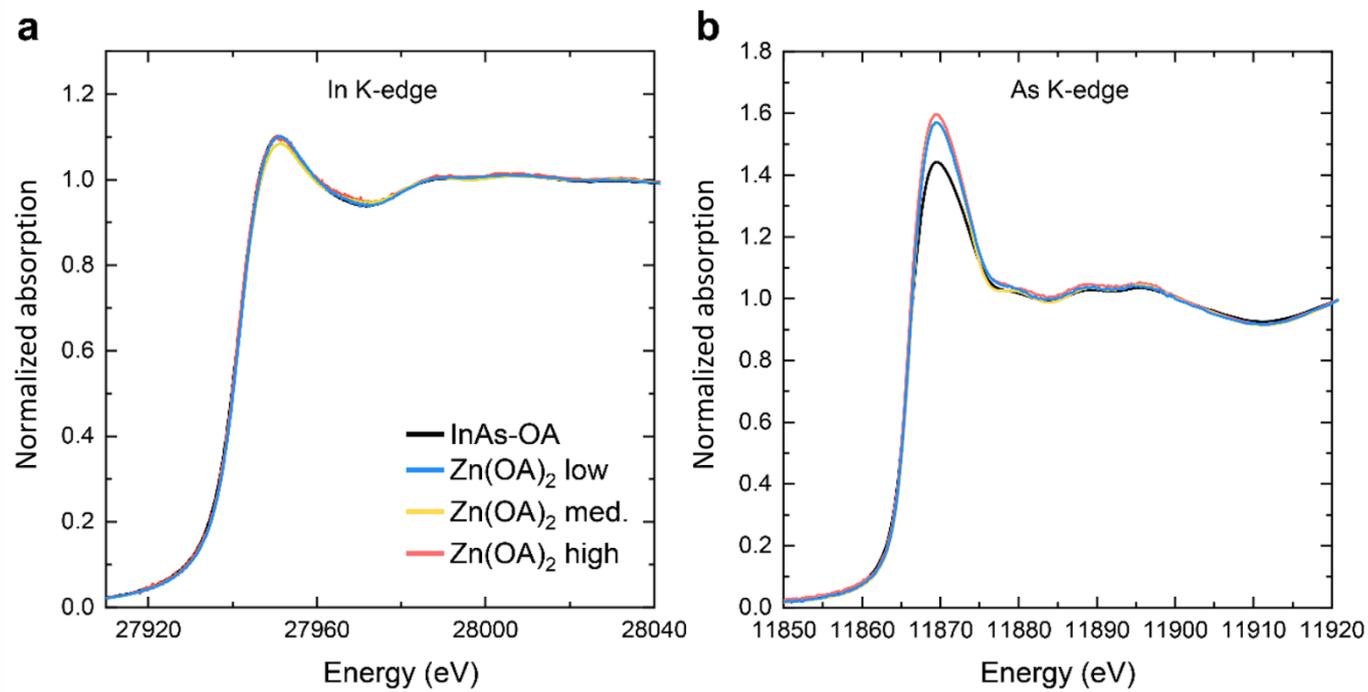

Figure S7. In K-edge (a), and As K-edge (b) normalized XANES spectra of Zn(Oleate)$_2$ doped InAs samples

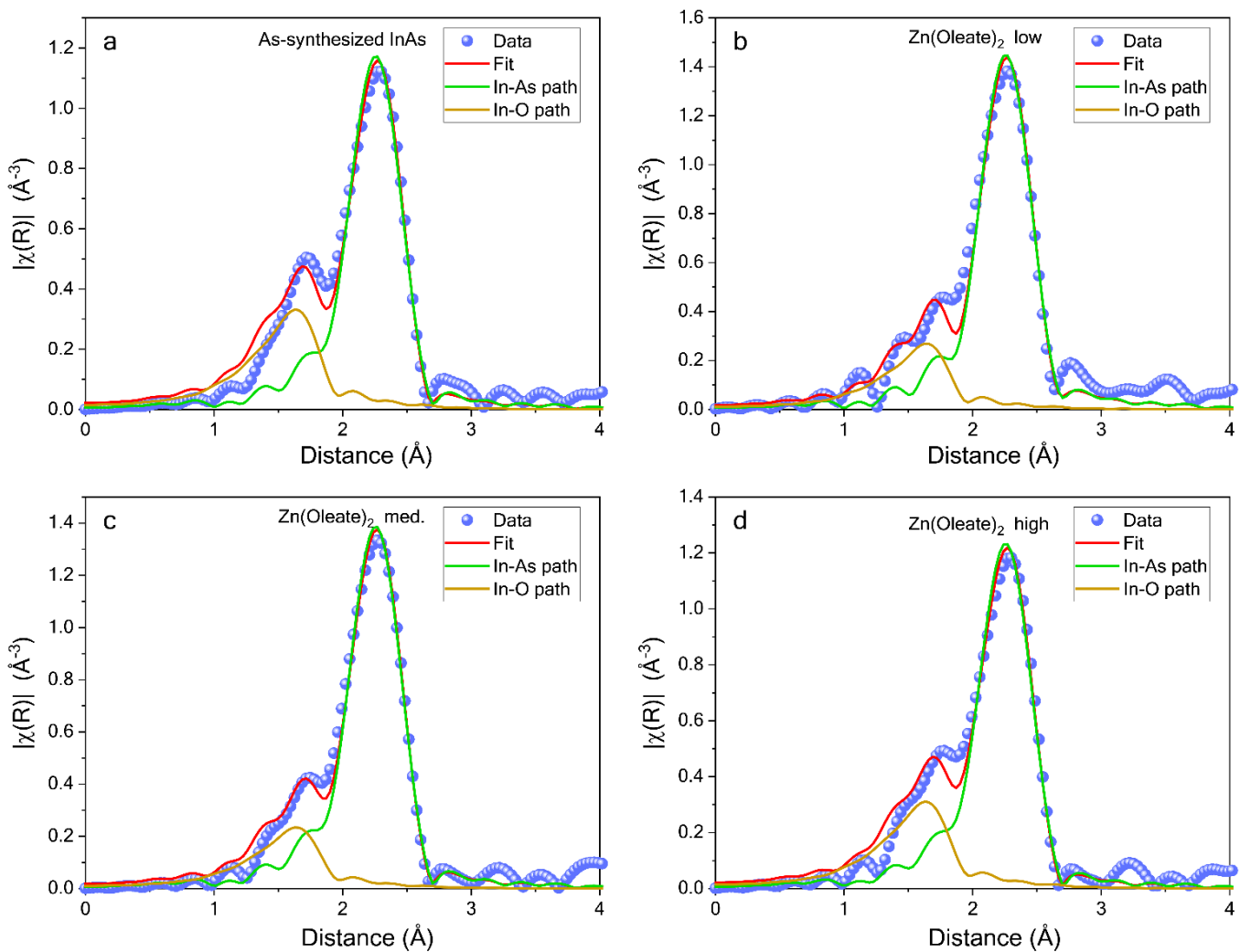

Figure S8. Fourier transform magnitude of In k$^2$-weighted EXAFS data of as-synthesized and Zn(Oleate)$_2$-doped InAs QDs samples (k-range: 2–14 Å$^{-1}$; rbkg: 1.3). Data (blue spheres) and fit (red line) to a structural model of Indium with two scattering paths of In-O and In-As (dark yellow and green lines, respectively)

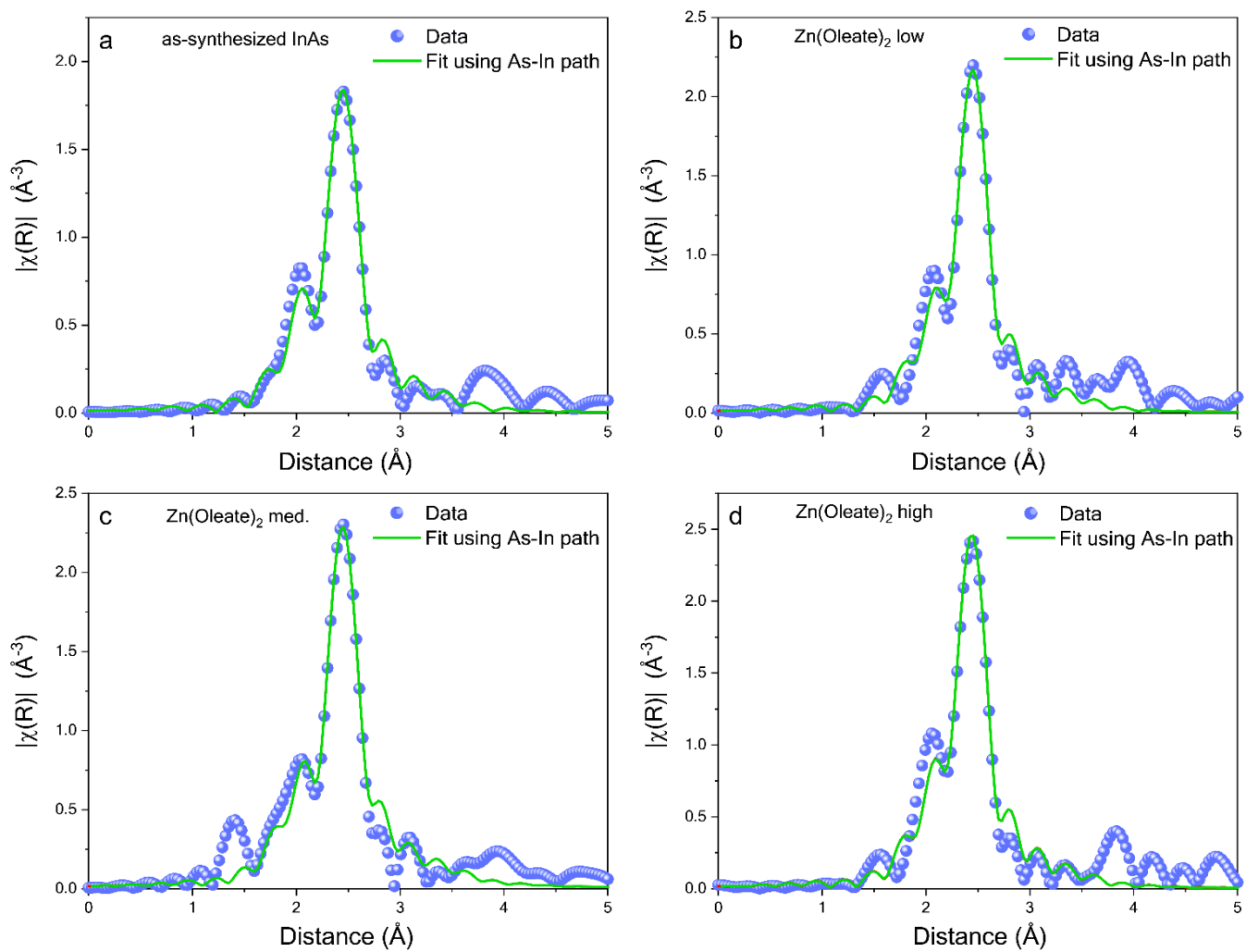

Figure S9. Fourier transform magnitude of As k[2]-weighted EXAFS data of as-synthesized and Zn(Oleate)$_2$-doped InAs QDs samples (k-range: 2–12 Å$^{-1}$; rbkg: 1.7). Data (blue spheres) and fit (red line) to a structural model of Indium with one scattering paths of As-In (green lines)

**Table S2.** Fitted parameters for EXAFS analyses of In, As and Zn edges from Zn(Oleate)$_2$-doped InAs NCs samples

| Sample number | Sample Name | Edge | Bond | C.N. | R(Å) | $\sigma^2$ (Å$^2$) | $\Delta E_0$ (eV) |
|---|---|---|---|---|---|---|---|
| 1 | InAs | As | As-In | 4 (fixed) | 2.609(4) | 0.0043(05) | 3.6(6) |
| | | In | In-As | 3.6(4) | 2.612(3) | 0.0047(07) | 4.3(5) |
| | | | In-O | 1.2(2) | 2.126(16) | 0.0052(29) | |
| 8 | Zn(Oleate)$_2$ X3 | As | As-In | 4.7(6) | 2.61(1) | 0.0037(06) | 3.7(1.0) |
| | | In | In-As | 3.0(1.0) | 2.608(9) | 0.0036(22) | 4.3(4) |
| | | | In-O | 0.9(8) | 2.124(93) | 0.0162(450) | |
| | | Zn | Zn-O | 2.0(1) | 2.037(30) | 0.0047(125) | 11.0(3.0) |
| | | | Zn-As | 0.3(6) | 2.219(12) | | |
| 9 | Zn(Oleate)$_2$ X2 | As | As-In | 5.0(5) | 2.613(4) | 0.0044(05) | 4.1(7) |
| | | In | In-As | 4.1(3) | 2.612(2) | | 4.3(4) |
| | | | In-O | 0.9(1) | 2.134(19) | 0.0019(74) | |
| | | Zn | Zn-O | 2.7(7) | 2.014(27) | 0.0039(43) | 10.0(2.0) |
| 10 | Zn(Oleate)$_2$ X1 | As | As-In | 4.3(4) | 2.619(4) | 0.0039(05) | 4.0(8) |
| | | In | In-As | 3.9(6) | 2.611(4) | 0.0050(10) | 4.3(4) |
| | | | In-O | 1.2(3) | 2.127(28) | 0.0093(93) | |
| | | Zn | Zn-O | 2.6(6) | 1.989(19) | 0.0008(31) | 7.0(2.0) |

**Table S3.** Fitted parameters of In and As and Zn edges from Diethylzinc-doped InAs NCs samples kept under inert conditions

| Sample number | Sample Name | Edge | Bond | N | R(Å) | $\sigma^2$ (Å$^2$) | $\Delta E_0$ (eV) |
|---|---|---|---|---|---|---|---|
| 1 | InAs | As | As-In | 4 (defined) | 2.609(4) | 0.0043(05) | 3.6(6) |
|  |  | In | In-As | 3.6(4) | 2.612(3) | 0.0047(07) | 4.3(5) |
|  |  |  | In-O | 1.2(2) | 2.126(16) | 0.0052(29) |  |
| 2 | DEZn X3 A | As | As-In | 3.5(4) | 2.614(4) | 0.0039(06) | 3.9(8) |
|  |  | In | In-As | 3.3(2) | 2.614(3) | 0.0102(48) | 4.3(5) |
|  |  |  | In-O | 1.7(2) | 2.128(14) |  |  |
|  |  | Zn | Zn-O | 3.8(4) | 1.983(11) | 0.0056(16) | 5.9(1.1) |
| 3 | DEZn X3 B | As | As-In | 4.5(7) | 2.61(1) | 0.0045(08) | 4.0(1.2) |
|  |  | In | In-As | 4.0(4) | 2.608(4) | 0.0089(139) | 4.3(5) |
|  |  |  | In-O | 0.9(4) | 2.137(44) |  |  |
|  |  | Zn | Zn-O | 2.2(4) | 2.018(25) | 0.0001(03) | 10.0(2.0) |
|  |  |  | Zn-As | 0.5(2) | 2.157(24) |  |  |
| 4 | DEZn X3 C | As | As-In | 3.3(4) | 2.62(1) | 0.0050 (defined) | 4.3(1.6) |
|  |  | In | In-As | 3.8(4) | 2.610(3) | 0.0051(06) | 4.3(5) |
|  |  | Zn | Zn-O | 2.0(4) | 1.986(18) | 0.0001(24) | 7.0(2.0) |
|  |  |  | Zn-As | 0.26(08) | 2.123(26) |  |  |
| 5 | DEZn X2 | As | As-In | 3.3(3) | 2.61(1) | 0.0050 (defined) | 2.8(1.3) |
|  |  | In | In-As | 3.0(2.0) | 2.615(12) | 0.0055(30) | 4.3(4) |
|  |  |  | In-O | 2.0(2.0) | 2.113(45) | 0.0085(140) |  |
|  |  | Zn | Zn-O | 4.0(3) | 1.976(7) | 0.0058(10) | 5.3(7) |
| 6 | DEZn X1 A | As | As-In | 4.7(5) | 2.610(4) | 0.0043(06) | 4.1(8) |
|  |  | In | In-As | 4.0(3) | 2.609(5) | 0.0042(defined) | 4.3(4) |
|  |  | Zn | Zn-O | 3.0(4) | 2.007(19) | 0.0040(23) | 8.0(2.0) |
|  |  |  | Zn-As | 0.8(2) | 2.153(17) |  |  |
| 7 | DEZn X1 B | As | As-In | 3.3(5) | 2.62(1) | 0.0043(09) | 4.4(1.2) |
|  |  | In | In-As | 4.0(3) | 2.611(2) | 0.0047(04) | 4.3(5) |
|  |  | Zn | Zn-O | 3.3(2) | 1.973(8) | 0.0029(11) | 5.8(8) |
|  |  |  | Zn-As | 0.2(1) | 2.119(23) |  |  |

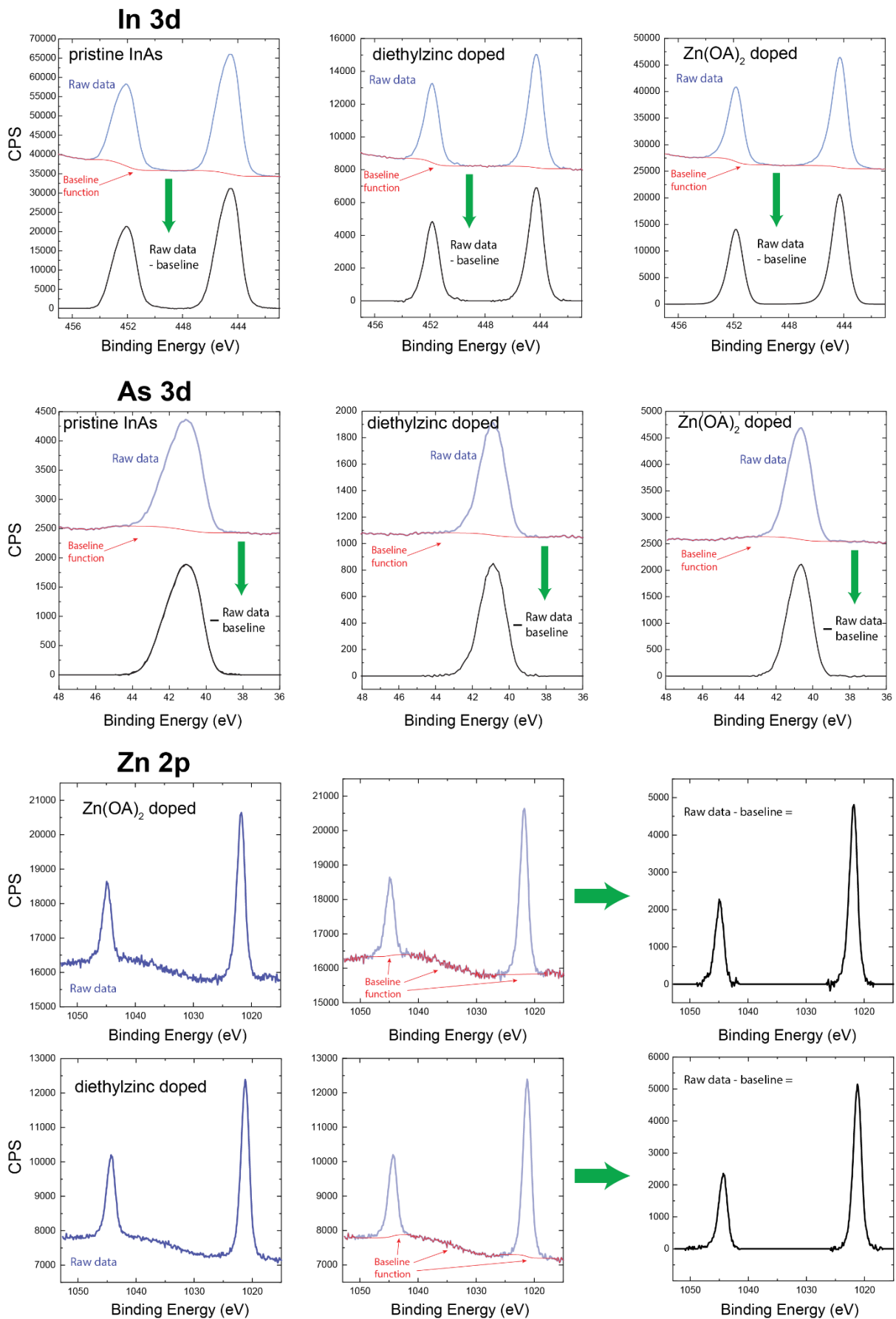

Figure S10. Raw data, baseline function and baseline subtraction of the XPS plots presented in figure 4 of the main manuscript.